\documentclass[12pt,a4paper,final]{iopart}
\usepackage[utf8]{inputenc}
\usepackage{graphicx,color}
\usepackage{dcolumn}
\usepackage{bm}
\usepackage{iopams}  
\usepackage{cleveref}
\usepackage{subfigure}
\usepackage[normalem]{ulem}
\usepackage{esvect} 

\usepackage{array}
\newcolumntype{L}{>{$}l<{$}} 
\usepackage{xcolor}

\begin{document}
\title{Thermal conductivity of porous polycrystalline PbTe}
\author{Javier F. Troncoso, Piotr Chudzinski, Tchavdar N. Todorov, Pablo Aguado-Puente, Myrta Gr\"uning and Jorge J. Kohanoff}
\address{Atomistic Simulation Centre, School of Mathematics and Physics, Queen's University Belfast, Belfast BT7 1NN, UK}%
\eads{\mailto{jfernandeztroncoso01@qub.ac.uk}}
\begin{abstract}
PbTe is a leading thermoelectric material at intermediate temperatures, largely thanks to its low lattice thermal conductivity. However, its efficiency is too low to compete with other forms of power generation. This efficiency can be effectively enhanced by designing nanostructures capable of scattering phonons over a wide range of length scales to reduce the lattice thermal conductivity. The presence of grain boundaries can reduce the thermal conductivity to $\sim 0.5$ Wm$^{-1}$K$^{-1}$ for small vacancy concentrations and grain sizes. However, grains anneal at finite temperature, and equilibrium and metastable grain size distributions determine the extent of the reduction in thermal conductivity. In the present work, we propose a phase-field model informed by molecular dynamics simulations to study the annealing process in PbTe and how it is affected by the presence of grain boundaries and voids. We find that the thermal conductivity of PbTe is reduced by up to 35\% in the porous material at low temperatures. We observe that a phase transition at a finite density of voids governs the kinetics of impeding grain growth by Zener pinning.
\end{abstract}


\maketitle


\section{Introduction}\label{sec:level1} 

The utilization of thermoelectric devices to convert heat into electric energy has been seen as a potential way of producing power since the discovery of the Seebeck effect in 1822. Thomas J. Seebeck observed that a temperature gradient between two electrical joints could produce a voltage difference. Twelve years later,  Jean Charles A. Peltier observed the reverse process in which a temperature gradient is created when a current flows between them. Thanks to these effects, thermoelectric materials can be used as power generators, thermoelectric coolers or optoelectronic devices. However, after nearly 200 years, the efficiency of currently available thermoelectric materials is still too low for most real-life applications as recently reviewed by Champier \cite{Champier2017} and, therefore, efforts are underway to improve the performance of these materials.

One of the most intensely investigated pathways to improve efficiency is the modification of the structure of the materials at the mesoscale. The thermodynamic efficiency of a thermoelectric material is proportional to its dimensionless figure of merit, $ZT=S^2 \sigma T/\kappa_T$, where $S$ is the Seebeck coefficient, $\sigma$ is the electrical conductivity, $T$ is the absolute temperature, and $\kappa_T$ is the total thermal conductivity, including electronic and phonon contributions. Only materials with $ZT > 1$ are considered suitable for practical applications at a given temperature. While all electronic properties are closely linked, the lattice thermal conductivity can be considerably reduced and plays an important role in the search for improved efficiency. Engineering the structure of the material at the micro and nanoscale (nanostructuring) can lessen the lattice conductivity considerably \cite{Hicks1993,Sutton1995,Dresselhaus2007,Zhao2014}. One way this can be achieved is by controlling the grain size in polycrystalline materials \cite{Yoneda2001,Kishimoto2002,Kuo2011,Yoon2013}. However, nanostructures might anneal at finite temperatures and distributions with small mean grain sizes may not be stable, which results in grain growth. The objective of the present study is to investigate this grain growth and the mechanisms by which the growth stops. We study the dependence of the mean grain size on temperature and vacancy concentration, considering the interaction between vacancies and grain boundaries as an essential factor.

The use of classical molecular dynamics (MD) simulations to study large structures is computationally demanding, so new models are required to analyze phenomena taking place at the mesoscale. In this work, we implement a phase-field method whose material-specific parameters are determined from MD simulations and energy-minimization calculations. For concreteness, we focus on one material: lead telluride (PbTe), one of the most widely studied and used thermoelectric material at intermediate temperatures, largely thanks to its low lattice thermal conductivity, with $\kappa\approx 2$ Wm$^{-1}$K$^{-1}$ at $300$ K. This thermal conductivity can be reduced in the presence of vacancies and grain boundaries, which can separately bring this value down to $\sim 0.5$ Wm$^{-1}$K$^{-1}$ \cite{JavPabJor2019}. In the regimes of high densities of point or planar defects, this conductivity becomes practically independent of temperature, as predicted by phenomenological models. We observed that grains grow until they reach a temperature-independent limiting grain size which depends on the concentration and size of voids. Known that grain boundaries can be stabilized by voids, and given that grain boundaries and voids are phonon scattering centers, we observed that porous polycrystalline samples can reduce the thermal conductivity of PbTe by up to a $35$ \%, and therefore enhance the thermoelectric efficiency.

In Section \ref{sec:methods}, the phase-field model is described to study the diffusion of the vacancy concentration and the time evolution of grain boundaries in PbTe. In Section \ref{sec:results}, we present our results. We first study grain growth in the presence of immobile voids of fixed size, and identify a kinetic phase transition as a function of void fraction. We then analyze the metastability of voids in the absence of grain boundaries and, by means of a thermodynamic model, we obtain their most probable size as a function of temperature. In Section \ref{ssec:c5_zenerpinning} we combine these two aspects to study the effect of voids as pinning particles for grain boundary stabilization, showing that a large number of small voids can stabilize small grains with low thermal conductivity. Finally, we use these results to obtain the thermal conductivity of porous polycrystalline PbTe as a function of temperature. In Sections \ref{sec:dicussion} and \ref{sec:conclusions} we present a discussion of these results and elaborate our conclusions, respectively.

\section{Phase-field modeling}\label{sec:methods} 
\begin{figure}[h!]
\centering
\subfigure[]{%
\includegraphics[width=0.33\columnwidth]{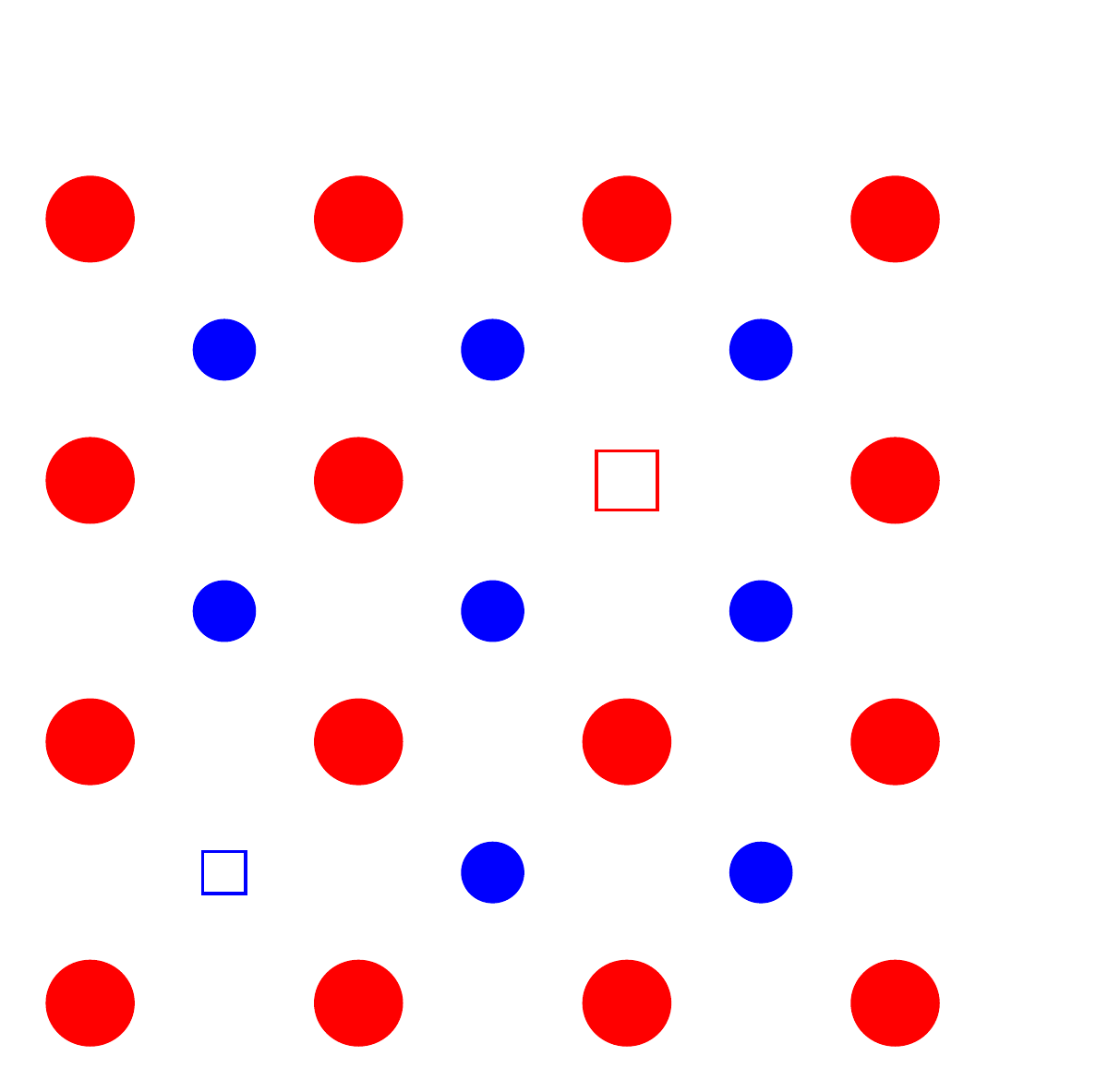}\hspace{2cm}
\label{fig:schottkya}}
\quad
\subfigure[]{%
\includegraphics[width=0.33\columnwidth]{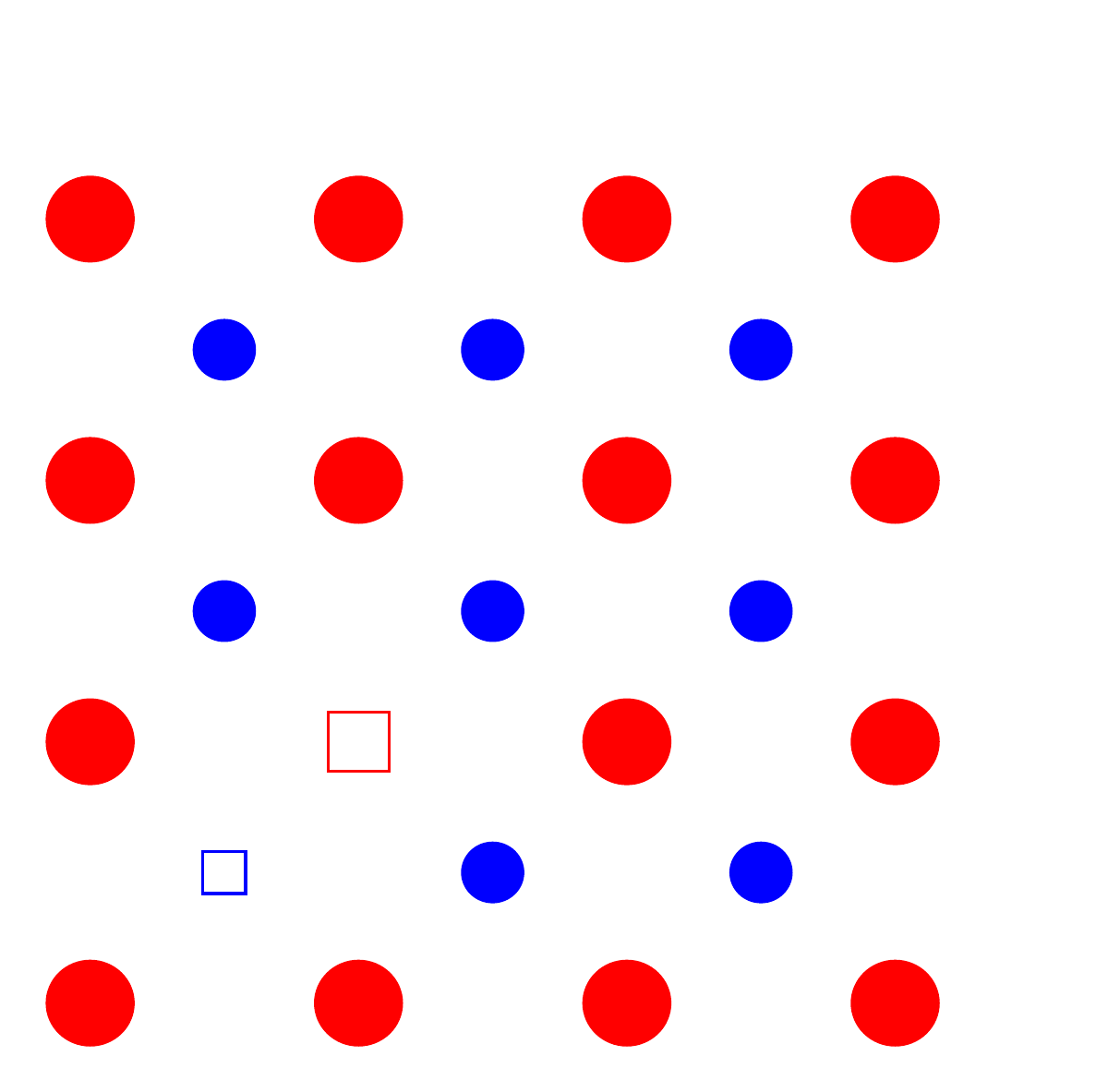}
\label{fig:schottkyb}}
\caption{ (a) Schottky pair, formed by a pair of isolated vacancies, and (b) Schottky dimer, formed by 2 consecutive vacancies, i.e., a divacancy. The formation energy of a Schottky dimer is smaller, and therefore this defect is more favorable. Circles stand for atoms while rectangles stand for vacancies (red for Pb and blue for Te).} 
\label{fig:schottky}
\end{figure}

A phase-field model is a method to study the diffusion of components and the dynamics of interfaces at the mesoscale. In this model, a system with interfaces and concentration gradients evolves in such a way that the contact surface between phases is minimized and the concentration tends to adopt the equilibrium value. We develop a phase-field model to describe the interaction between vacancies and grain boundaries in polycrystalline PbTe. We consider the equations of motion for the concentration of Schottky defects (see Fig. \ref{fig:schottky}) \cite{Cahn1958}, $c_v$, and for order parameters \cite{Allen1979}, $\eta_\alpha$, describing $N$ grain orientations. There are as many order parameters as grain orientations, $N$, and they take the value $1$ inside the grain that they represent and $0$ outside, with a value between 0 and 1 at the grain boundary.  If the number of orientations is smaller than the number of order parameters, $N$ has to be large enough to ensure that two grains with the same orientation do not come into contact \cite{Kim2006,Krill2002}. For simplicity, we only work with systems where all grains have different orientations.

According to the phase-field method, the temporal evolution of the vacancy concentration, $c_v$, is governed by the Cahn-Hilliard diffusion equation \cite{Cahn1958}: 

\begin{equation}
\frac{\partial c_v}{\partial t}=\vec{\nabla}\left(M\vec{\nabla}\left(\frac{\partial f_{local}}{\partial c_v}-\kappa_v\nabla^2 c_v\right)\right)+S_v+\zeta_c(T),
\label{eq:cahnhilliard}
\end{equation}
and the evolution of local order parameters, $\eta_\alpha$, is described by the Allen-Cahn equation \cite{Allen1979}:
\begin{equation}
\frac{\partial\eta_\alpha}{\partial t}=-\frac{L}{V_m}\left(\frac{\partial f_{local}}{\partial \eta_\alpha}-\gamma\nabla^2\eta_\alpha\right)+\zeta_\eta(T),
\label{eq:allencahn}
\end{equation}
where $M$ is the vacancy mobility, $L$ is the interface mobility, $V_m$ is the molar volume, $f_{local}$ is the local free energy per mole, $T$ is the absolute temperature, $\kappa_v$ and $\gamma$ are gradient energies and $\zeta$ are Gaussian noises that reproduce thermal fluctuations in the system.  These fluctuations will be described explicitly in Section \ref{sec:thfluct}. In the isotropic case, $M$ and $L$ are constant, while they depend on the interacting grains in the anisotropic material. The term $S_v$ accounts for the generation or annihilation rates of vacancies and will be discussed in Section \ref{sec:recombinationrates}. 
Equation \ref{eq:cahnhilliard} is equivalent to Fick's law for the diffusion of components and the total number of vacancies in the system is conserved if the last two terms, i.e. the production/annihilation term and the Gaussian noise, are neglected. 
The vacancy mobility, $M$, plays an essential role in the diffusion process and will be discussed in Section \ref{sec:diffcoef}.

On the other hand, Eq. \ref{eq:allencahn} is applied to the order parameters and describes the phase separation in a system with concentration gradients. The order parameters $\eta_\alpha$ are not conserved due to the fact that individual grains are allowed to expand, shrink and also disappear, and, therefore, the number of grains is not constant. These order parameters change only at  grain boundaries and these grain boundaries move to reduce the total free energy of the system. The grain boundary velocity depends on the interface mobility, $L$, which will be described in Section \ref{sec:gbparams}. This equation is solved for all order parameters and the following constraint has to be fulfilled at each point in space:
\begin{equation}
\sum_\alpha \eta_\alpha=1\quad\textnormal{    with } \eta_\alpha  \in [0,1]\quad \forall \alpha.
\label{eq:orderparamsconstraint}
\end{equation}
In practice, this constraint is imposed by renormalising the amplitude of the phases after the integration step \cite{Kim2006}:
\begin{equation}
\eta_\alpha^\textnormal{a}=\frac{\eta_\alpha^\textnormal{b}}{\sum_\beta\eta_\beta^\textnormal{b}},
\label{eq:orderparamsconstraint}
\end{equation}
where the superscripts b and a stand for the order parameters before and after the reassignment respectively.

Equations \ref{eq:cahnhilliard} and \ref{eq:allencahn} are coupled since they are given as a function of the local free energy of the system, $f_{local}$. A good description of this free energy is therefore essential to capture the phase evolution accurately. Our approximation for $f_{local}$ is given in Section \ref{sec:freeenergy}.

\subsection{Free energy}\label{sec:freeenergy} 
The local free energy density, $f_{local}$, in Eqs.  \ref{eq:cahnhilliard} and \ref{eq:allencahn} is the sum of the free energy of the homogeneous material, the excess free energy due to the presence of grain boundaries in the material, and additional contributions, such as the elastic free energies or external fields (which are not considered in this work). We propose a simple model where the local free energy is the energy inside individual grains supplemented with the excess free energy at the grain boundary:
\begin{eqnarray}
f_{local}=  f(c_v,T)\sum_\alpha\eta_\alpha^2+\sum_{\alpha\neq \beta}W_{\alpha\beta}\eta_\alpha\eta_\beta,
\label{eq:freeen}
\end{eqnarray}
\noindent with
\begin{eqnarray}
f(c_v,T)&=&h_vc_v+f_2c_v^2+f_3c_v^3+f_4c_v^4 \nonumber \\
&+&RT\left[c_v\textnormal{ ln }c_v+(1-c_v)\textnormal{ ln }(1-c_v)\right],
\label{eq:freeenergy}
\end{eqnarray}

\noindent where $W_{\alpha\beta}$ is the excess free energy between grains $\alpha$ and $\beta$, $h_v$ is the formation enthalpy and $f_2,f_3,f_4$ are mixing terms such that a void-free system and a system with voids are equally stable. The second term in Eq. \ref{eq:freeen} corresponds to the energy cost necessary to move from phase $\alpha$ to phase $\beta$ and $W_{\alpha\beta}$ is the energy cost. We will see that this energy depends on the misorientation angle, $\theta$, between grains $\alpha$ and $\beta$,  $W_{\alpha\beta}\equiv W(\theta)$, and that it is related to the grain boundary energy (see Section \ref{sec:gbparams}). This misorientation angle is the relative angle between the lattice vectors of two adjacent grains.  According to the second term in Eq. \ref{eq:freeen}, the lowest free energy occurs away from the grain boundaries, where the product $\eta_\alpha\eta_\beta$ is equal to zero, and grain boundaries evolve to remove this excess grain boundary energy. In the presence of vacancy concentrations different from equilibrium values, the first term in Eq. \ref{eq:freeen} increases and grain growth is slowed down so that grain boundaries can absorb the excess of vacancies, according to the production/annhilation term $S_v$ in Eq. \ref{eq:cahnhilliard} that we will describe below. This first term in Eq. \ref{eq:freeen} falls at the grain boundary, where the vacancy formation energy decreases \cite{Tschopp2012}.

\begin{figure}
        \centering
        \includegraphics[width=0.7\textwidth]{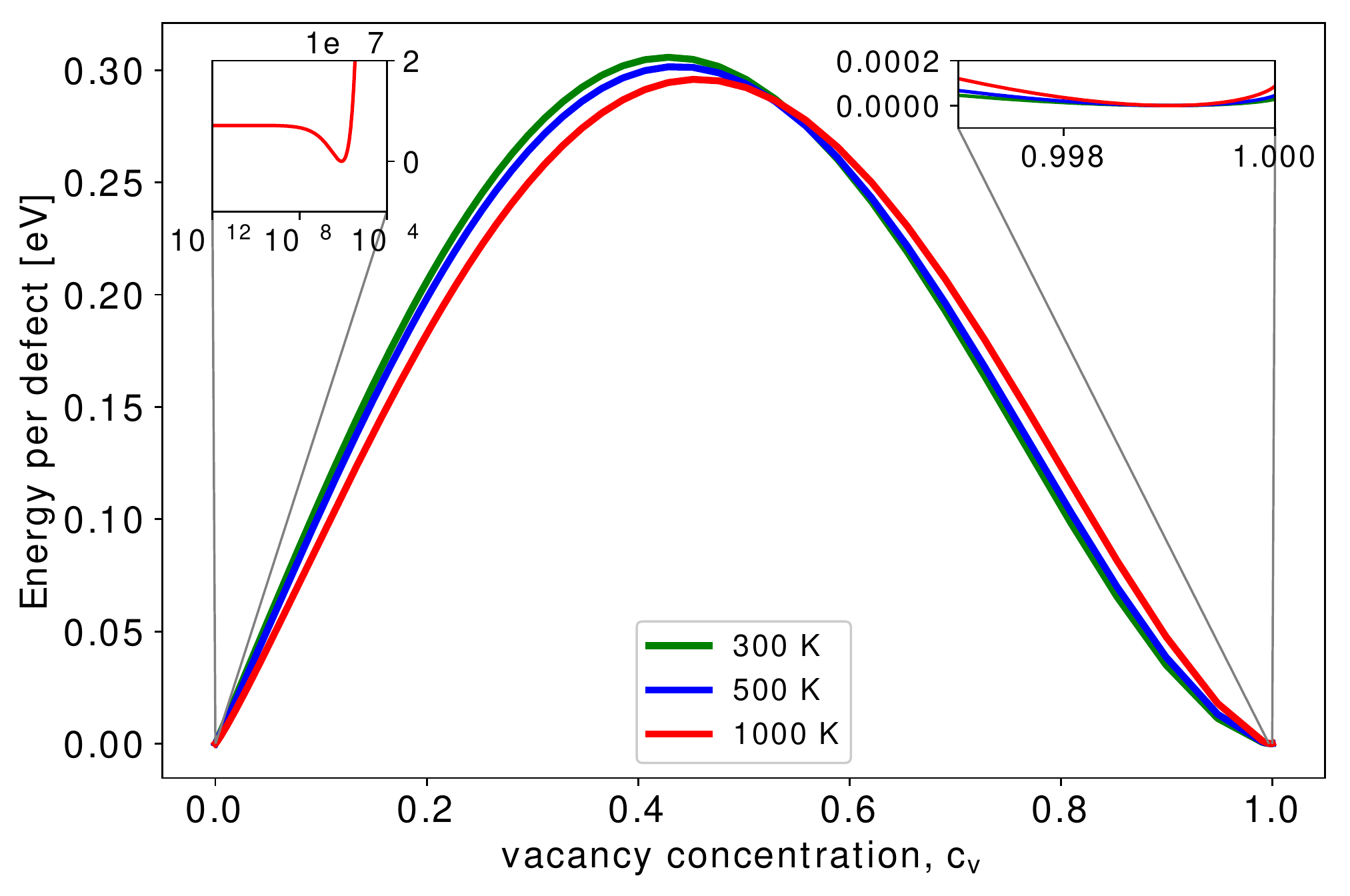} 
        \caption{\small Chemical free energy of a system with vacancies and voids. 
        }
        \label{fig:finalfreeen}
\end{figure}

The free energy defined in Eq. \ref{eq:freeenergy} corresponds to the bulk free energy of the system in the absence of grain boundaries. The model parameters $h_v,f_2,f_3$ and $f_4$ are determined by enforcing the following conditions: a void and a region in which the vacancy concentration is in equilibrium are equally stable and their local free energy is equal to zero \cite{Li2011,Chang2016}. Voids are defined as structures where the vacancy concentration is equal to $0.999$. These conditions mean that the free energy and its first derivative are equal to zero when $c_v=0.999$ and $c_v=c_v^{\normalfont{eq}}$, where $c_v$ is the concentration of Schottky defects and $c_v^{\normalfont{eq}}$ is the equilibrium value taken from Ref. \cite{WunFanLi2015}. They result in the profile observed in Fig. \ref{fig:finalfreeen}, where the insets show the behavior close to the minima.

The presence of vacancies and grain boundaries can also give rise to additional stresses and strains leading to an elastic energy contribution. However, the calculation of this free energy increases considerably the simulation times and it does not play a relevant role under small defect concentrations or in the absence of external stresses. Under normal conditions, the estimated driving pressure associated with grain boundary motion due to the excess grain boundary energy is higher than the driving pressure due to the elastic energy by two orders of magnitude \cite{Gottstein2002}.

The gradient coefficient for vacancies, $\kappa_v$, is linked to the mixing energy according to $\kappa_v=f_2l_v^ 2/2$, where $l_v$ is the typical distance between vacancies.

\subsection{Interface mobility and energy barrier}\label{sec:gbparams}
The interface mobility, $L$, in Eq. \ref{eq:allencahn} is associated with the grain boundary velocity and proportional to the grain boundary mobility, $m$:  
\begin{equation}
L=\frac{\pi^2m}{8\Delta_{gb}}
\label{eq:interfacemobilitymodel}
\end{equation}
where $\Delta_{gb}$ is the grain boundary thickness. We use $\Delta_{gb}=6\Delta x$, where $\Delta x$ is the grid spacing, according to convergence tests performed to minimize the error in the steady-state boundary velocity and optimize the computational accuracy and cost \cite{Miyoshi2016}.  Similarly, the parameter $W$ in equation \ref{eq:freeen} is closely linked to the grain boundary energy, $\sigma$, given in J m$^{-2}$. The free energy, and therefore $W$, is given in J mol$^{-1}$, so the next equation is extracted:
\begin{equation}
W=\frac{4\sigma V_m}{\Delta_{gb}},
\label{eq:interfaceenergymodel}
\end{equation}
with $V_m$ the molar volume.
In the anisotropic case, the grain boundary mobility and energy are given as a function of the misorientation angle, as we will see below. 

The gradient-free energy defined in Eq. \ref{eq:allencahn} requires the use of a gradient coefficient, $\gamma$. If one considers the free energy as a whole, it is easy to see that $W$ and $\gamma$ have to be linked in order to have stable simulations and to prevent the growth of third phases in the interfaces \cite{Steinbach2009}. This relation is given through the grain boundary energy \cite{Kim2006}:
\begin{equation}
\gamma=\frac{8\sigma \Delta_{gb} V_m}{\pi^2}=\frac{2 W\Delta_{gb}^2 }{ \pi^2}
\label{eq:interfacegrademodel}
\end{equation}

The grain boundary energy and mobility were obtained from energy-minimization calculations and MD simulations, respectively. The grain boundary energy, $\sigma$, is calculated from the excess free energy due to the presence of one grain boundary as
\begin{equation}
\sigma = \frac{E - n_aE^0}{A},
\label{eqn:gbe}
\end{equation}
where $E$ is the total energy, $E^0$ is the energy per atom in the bulk material, $A$ is grain boundary area, and $n_a$ is the number of atoms in the simulation cell. 

The calculation of grain boundary mobilities requires the study of the grain boundary velocity during grain boundary motion as a function of the driving pressure. For this study, we ran MD simulations using a bicrystal, the force field of Ref. \cite{JavPabJor2019}, and an additional artificial potential,
$u_\xi$, applied to the atoms of one of the grains forming the simulation box. This artificial potential corresponds to predefined forces added to the atoms at the grain boundary to favor the growth of one grain. Although this potential has no physical meaning, it has been proved that the mobilities do not depend on the driving force \cite{Janssens2006,Sutton1995}. In a bicrystal formed by two grains where one of them presents a higher free energy per atom,  the interface always moves towards the grain with the higher free energy to reduce the overall free energy of the system \cite{Sutton1995}. The artificial potential used to provide the driving force has been proposed by Janssens \cite{Janssens2006} and later adapted in the LAMMPS code \cite{Plimpton1995,lammpsWeb} to describe a lattice with 6, instead of 12, first neighbors. The potential energy added to each atom in one of the grains is given by:
\begin{equation}
u_{\xi}(\vec{r}_i) =
\left\{ \begin{array}{@{\kern2.5pt}lL}
     0 & if $\xi_i<\xi_l$.\\
    \hfill 0.5V(1-\cos(2\omega_i)) & if $\xi_l<\xi_i<\xi_h$.\\
     V & if $\xi_h<\xi_i$.
\end{array}\right.
\label{eq:artpot}
\end{equation}

where $\xi_i$ is the order parameter of each atom $i$ given by a sum over its six first neighbors: 
\begin{equation}
\xi_i=\sum_j^6 |\vec{r}_j-\vec{r}_j^I|.
\end{equation}
Here, $\vec{r}_j^I$ is the nearest ideal lattice site of crystallite $I$ to $\vec{r}_j$. Crystallite $I$ is the grain whose atoms will not experience the additional potential. Atoms with $\xi_i<\xi_l$ belong to this grain, atoms with $\xi_h<\xi_i$ belong to the other grain and receive the extra potential energy, and atoms with intermediate values are at the grain boundary and are subject to an artificial force. The limits $\xi_h$ and $\xi_l$ take the values $0.25$ and $0.75$, respectively, to ensure that forces are only added to atoms located at the grain boundary. The parameter $\omega_i$ is defined as
\begin{equation}
\omega_i=\frac{\pi}{2}\left(\frac{\xi_i-\xi_l}{\xi_h-\xi_l}\right).
\end{equation}

According to this definition, the atoms of one of the grains experience an additional potential that goes to zero once they migrate into the other grain. The parameter $V$ determines the energy per atom added associated with a specific element type \cite{Janssens2006} and is set to $V=1$ kcal/mol per atom. The bicrystal is equilibrated without the artificial potential for 250 ps and then the mobility is calculated from the grain boundary velocity once the artificial potential is applied, between 200 ps and 1 ns, depending on the relative misorientation.
 
\begin{figure}[h!]
        \centering
        \includegraphics[width=0.7\textwidth]{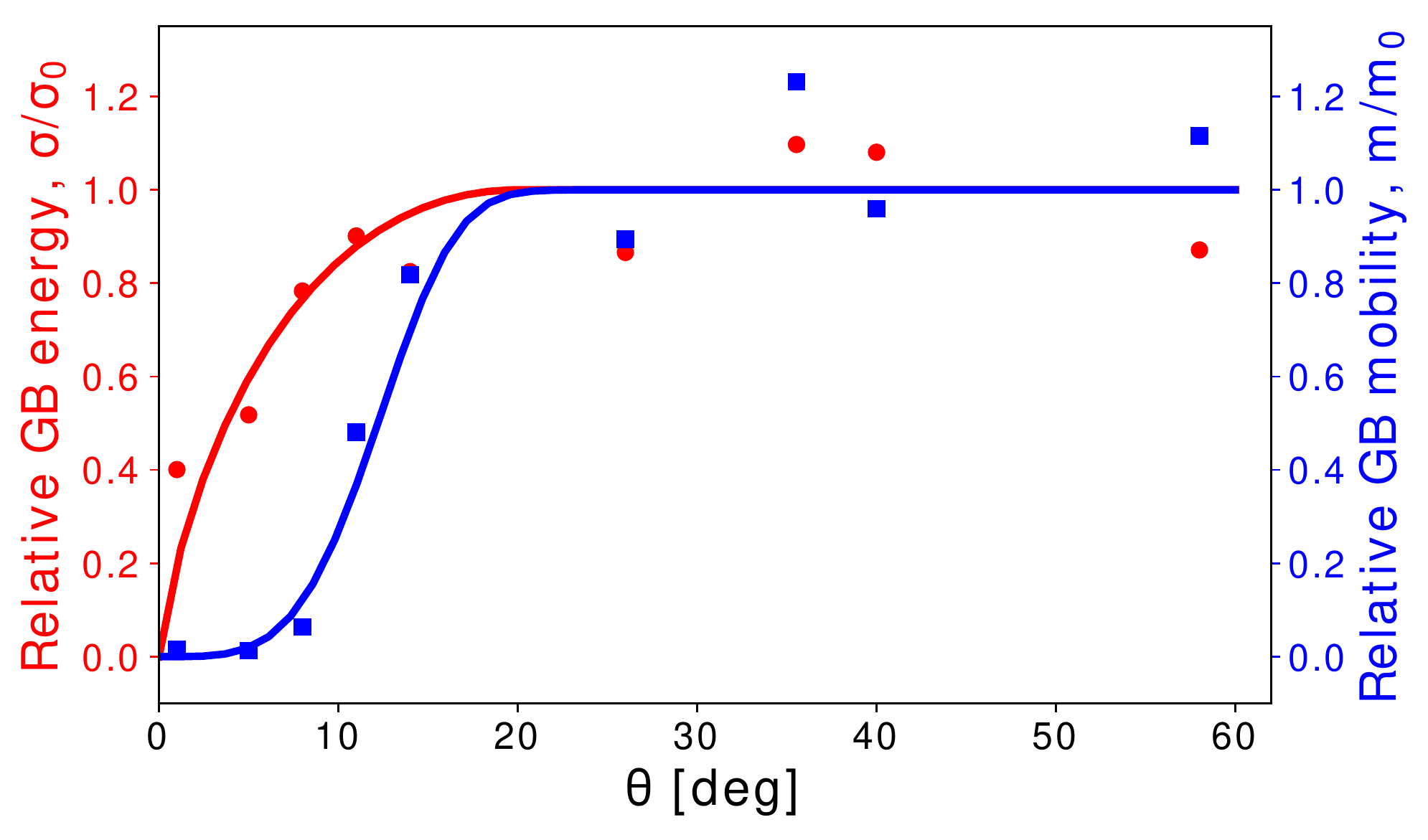} 
        \caption{\small Grain boundary energy and mobility as a function of the misorientation angle in a $\langle 100\rangle$ tilt boundary. Red dots represent the relative grain boundary energy and were calculated using Eq. \ref{eqn:gbe} in a bicrystal containing 2 grains with different misorientation angles. The red line corresponds to Eq. \ref{eq:gbenergy} with $\theta_m=20^{\circ}$ and $\sigma_0=2410$ mJ m$^{-2}$. The blue line corresponds to Eq. \ref{eq:gbmobility} with $\theta_m=20^{\circ}$ $m_0=1.5$ m s$^{-1}$ MPa$^{-1}$, and $q_m=0.027$ eV. The relative grain boundary mobility as a function of the misorientation is plotted with blue squares. Results for the mobility were obtained from MD simulations using the artificial potential described by Eq. \ref{eq:artpot} and are compared with the model described by Eq. \ref{eq:gbmobility} (blue line). These results show a distinction between low-angle and high-angle grain boundaries \cite{Sutton1995}.}
        \label{fig:gbfrommd}
\end{figure}

The grain boundary energy and mobility calculated in LAMMPS were compared with phenomenological models proposed in the literature. The energy of one grain boundary between two grains with relative misorientation $\theta$ is approximated as \cite{Read1950,Read1953}:

\begin{equation}
\sigma(\theta) =
\left\{ \begin{array}{@{\kern2.5pt}lL}
    \hfill \sigma_0\left(\theta/\theta_m\right)\left(1-\textnormal{ln}\left(\frac{\theta}{\theta_m}\right)\right) & if $\theta<\theta_m$.\\
    \sigma_0 & if $\theta>\theta_m$.
\end{array}\right.
\label{eq:gbenergy}
\end{equation}
where the parameters $\sigma_0$ and $\theta_m$ are independent of misorientation and temperature. This equation is obtained directly from Dislocation Theory \cite{Read1953}, where a grain boundary is considered as an array of dislocations. This energy saturates to $\sigma_0$ at large angles, except for special angles in which the lattices match (coincidence site lattices, CSLs \cite{Read1953,Sutton1995,Gottstein2002,Bulatov2014}). For simplicity, these cases are not taken into account in the present study. $\theta_m$ is the angle at which the grain boundary energy becomes constant and equal to $\sigma_0$. It is obtained, together with $\sigma_0$, by fitting the data reported in Fig. \ref{fig:gbfrommd} (red circles), with Eq. \ref{eq:gbenergy}. The fitted values are $\sigma_0=2410$ mJ m$^{-2}$ and $\theta_m=20^{\circ}$. The effect of the inclination angle, i.e. the angle between the grain boundary and the plane perpendicular to the misorientation axis between adjacent grains, on grain growth is negligible in comparison with the effect of the misorientation angle \cite{Shahnooshi2019}. Therefore, $\sigma_0$ is often considered also independent of inclination \cite{Shahnooshi2019,Miyoshi2017}. 

Similarly, the mobility of a grain boundary also depends on the misorientation angle \cite{Humpreys1997}:
\begin{equation}
m(\theta,T)=
         \bar{m}(T)\left(1- e^{-5\left(\theta/\theta_m\right)^4}\right),
\label{eq:gbmobility}
\end{equation}
where $\bar{m}(T)=m_0\exp(-q_m/k_BT)$ is temperature-dependent. The temperature-independent mobility, $m_0$, and the corresponding activation energy, $q_m$, are obtained by fitting $\bar{m}(T)$ for temperatures in the range 300-1000 K. The fitted values are $m_0=1.5$ m s$^{-1}$ MPa$^{-1}$ and $q_m=0.027$ eV. These models are in good agreement with our results, as can be seen in Fig. \ref{fig:gbfrommd}. While the dependence of mobility and energy on the misorientation angle is important for small angles, the dependence on the misorientation axis is small. Therefore, in practice, we use the same value, averaged over bicrystals with different misorientation axes, for all axes. 

\subsection{Diffusion coefficient}\label{sec:diffcoef} 
The parameter $M$ in Eq. \ref{eq:cahnhilliard} is the vacancy mobility and is associated with the vacancy velocity in bulk PbTe. $M$ is related to the diffusion coefficient $D_v$ by the Nernst-Einstein relation:
\begin{equation}
M=\frac{D_v c_v}{RT}.
\end{equation}

The diffusion coefficient was calculated from the mean square displacement of atoms in classical MD simulations, in the presence of a Schottky defect \cite{Blochl1993,Hu2002},
\begin{equation}
D_v=\frac{1}{6nt}\sum^n_i \left(\vec{r}_i(t)-\vec{r}_i(0)\right)^2.
\label{eq:diffcoeff}
\end{equation}
In an MD simulation at constant temperature, the diffusion coefficient is calculated from the mean square displacements of the closest atom to each of the $n$ vacancies in the simulation box for a time $t$. $\vec{r}_i(t)$ is the position of the atom $i$ at time $t$. All MD simulations performed in the present study were run using the force field of Ref. \cite{JavPabJor2019}. These simulations were run for up to $10$ ns, with the longest simulations at lower temperatures, where the number of vacancy jumps is smaller, as in this case the diffusion coefficient may be overestimated. Nevertheless, this would not have a strong impact on grain growth due to the high production of vacancies due to external sources. The diffusion coefficient is calculated at high and intermediate temperatures. 

In the study of grain growth in porous polycrystalline PbTe we will consider voids, which are regions of very high vacancy concentration. In Sections \ref{sec:mgs} and \ref{ssec:c5_zenerpinning}, voids will be taken as immobile, and hence the diffusion constant of vacancies belonging to these voids will be $D_v=0$. The size of these voids will be fixed. On the other hand, in Section \ref{sec:voids}, voids will be allowed to evolve according Eq. \ref{eq:cahnhilliard} to study void growth. In practice, voids block grain boundary motion and grain boundaries continue their movement by surrounding them.

\subsection{Grain boundaries as vacancy sources/sinks}\label{sec:recombinationrates} 
Grain boundaries can act as vacancy sources/sinks in the presence of radiation or external stress \cite{Humphreys2004,Tschopp2012,Lejcek2010}.
Moreover, grain boundaries act as vacancy sinks during quenching, and therefore the vacancy concentration falls at grain boundaries \cite{Faulkner2009}. However, although these effects are observed, the way how defects interact with grain boundaries is not fully understood \cite{Bai2010}.

The production/annihilation of vacancies at the grain boundary is modelled by the term $S_v$ in Eq. \ref{eq:cahnhilliard} as follows:
\begin{equation}
S_v^{GB}=-s^{GB}_v(c_v-c_v^{eq})\left(1-\sum_{\alpha}\eta_\alpha^2\right),
\label{eq:recombinationrates}
\end{equation}
where $s^{GB}_v$ is the sink strength in [t$^{-1}$] units, meaning that it is related to the relaxation time, i.e. $s^{GB}_v$ is the inverse of the time required by the vacancy concentration to drop to the equilibrium concentration at the grain boundary. This relaxation time is in the range of 1-50 ns \cite{Millett2009}, and the dependence on the grain boundary type (continuous - low angle or sharp - high angle) is neglected. We observed that this relaxation time affects how fast equilibrium is reached, but not the final configuration. The model in Eq.  \ref{eq:recombinationrates} is defined such that the production/annihilation of vacancies occurs at grain boundaries but not far from them, where the term into parenthesis falls to zero. According to the model in Eq. \ref{eq:recombinationrates}, grain boundaries act as vacancy sinks in over-saturated systems, where $c_v>c_v^{eq}$, and as vacancy sources in under-saturated systems, where $c_v<c_v^{eq}$.

A system with a vacancy concentration far from the equilibrium value tends to adopt the equilibrium value. Vacancies would tend to recombine with interstitials and reach the equilibrium values after a long - and unknown - time. This mechanism would depend on the concentration of both interstitials, $c_i$, and vacancies as $c_v\cdot c_i$. However, the diffusion coefficients of vacancies are orders of magnitude larger than those of interstitials \cite{Gilmer1995}, so vacancies and interstitials produced simultaneously are annihilated at the grain boundaries at different times. The formation energy of interstitials is higher than that of vacancies \cite{Li2015}, so $c_i$ is usually much smaller than the vacancy concentration. In the present work, only the interaction between vacancies and grain boundaries is studied, so $c_i$ is not considered. We are interested in systems with high vacancy concentrations since their lattice thermal conductivity falls considerably. Therefore, the vacancy concentration inside individual grains is set roughly constant. This effect can be produced by continuous irradiation and enters into Eq. \ref{eq:cahnhilliard} as follows:

\begin{equation}
S_v^{G}=-s^{G}_v(c_v-c_v^{0})\left(\sum_{\alpha}\eta_\alpha^2\right),
\label{eq:recombinationratespluskeephigh}
\end{equation}
where $c_v^0$ is the initial vacancy concentration and $s^{G}_v$ is the inverse relaxation time, which is set to $s_v^{G}=1$ ns$^{-1}$, and vacancies are loaded into the grains, especially. $c_v^0$ is the concentration that the system tends to adopt in the presence of external sources,  e.g. radiation. In their absence, this value corresponds to the equilibrium vacancy concentration. The term $S_v$ in Eq. \ref{eq:cahnhilliard} corresponds to $S_v=S_v^{GB}+S_v^{G}$.

\subsection{Thermal fluctuations}\label{sec:thfluct}
At finite temperature vacancies and grain boundaries are subject to thermal fluctuations. These fluctuations can be represented in Eqs. \ref{eq:cahnhilliard} and \ref{eq:allencahn} by Langevin Gaussian noises which satisfy the fluctuation-dissipation theorem  \cite{Landau1980,Toth2010}:

\begin{equation}
\langle\zeta_c(T,\vec{r}_i,t)\rangle=0,
\label{eq:fluctconc1}
\end{equation}
\begin{equation}
\langle\zeta_c(T,\vec{r}_i,t)\cdot \zeta_c(T,\vec{r}_j,t')\rangle=2\frac{MRT}{\Delta t(\Delta x)^2}\delta(\vec{r}_i,\vec{r}_j)\delta(t,t'),
\label{eq:fluctconc2}
\end{equation}
\begin{equation}
\langle\zeta_\eta(T,\vec{r}_i,t)\rangle=0,
\label{eq:fluctph1}
\end{equation}
\begin{equation}
\langle\zeta_\eta(T,\vec{r}_i,t)\cdot \zeta_\eta(T,\vec{r}_j,t')\rangle=2\frac{Lk_BT}{\Delta t(\Delta x)^3}\delta(\vec{r}_i,\vec{r}_j)\delta(t,t'),
\label{eq:fluctph2}
\end{equation}
where $\Delta t$ and $\Delta x$ are the time step and grid spacing respectively and $\delta(\vec{r}_i,\vec{r}_j)$ is the Kronecker delta between grid points $\vec{r}_i$ and $\vec{r}_j$. Similarly, $\delta(t,t')$ corresponds to the discrete times $t$ and $t'$.

However, Eq. \ref{eq:allencahn} is only solved at the grain boundary and its vicinity. Additionally, we observed that the presence of thermal fluctuations does not have a strong impact on the grain boundary velocity, and therefore they can be ignored to speed up simulations.

\section{Results}\label{sec:results}

\subsection{Mean grain size distribution}\label{sec:mgs}
Equations \ref{eq:cahnhilliard} and \ref{eq:allencahn} were solved using the model parameters listed in Table \ref{table:properties} to study grain growth in porous polycrystalline PbTe and void stability. The three-dimensional simulation box was discretized in a Cartesian grid under periodic boundary conditions and these equations were solved at each grid point after each time step using finite differences and the forward Euler time integration method. A 27-point stencil for discrete Laplacian approximations is used \cite{OReilly2006}. In the study of grain growth, the grid spacing is set to $\Delta x=6.43$ nm and the simulation box is formed by a randomly generated polycrystalline structure which follows the Voronoi tessellation \cite{Aurenhammer1991} and a  distribution of equally spaced voids. The grid spacing is reduced to $\Delta x=6.43$ \AA~in the study of void stability in a simulation box containing one void at the center and in the absence of grain boundaries. These results will be shown in Section \ref{sec:voids}.

\begin{table}[h!]
\newcommand{\specialcell}[2][c]{%
  \begin{tabular}[#1]{@{}c@{}}#2\end{tabular}}
\newcolumntype{M}[1]{>{\centering\arraybackslash}m{#1}}
\newcolumntype{N}{@{}m{0pt}@{}}
\caption{\label{table:properties}PbTe properties used in the simulations. The statistical uncertainties of the diffusion coefficient and grain boundary mobility are 7\% and 2\%, respectively. The uncertainty in the grain boundary energy, $\sigma_0$, is negligible. The diffusion coefficient, $D$, and grain boundary mobility, $\bar{m}$, follow the Arrhenius law in temperature, $D_v(T)=D_0e^{-q_D/k_BT}$ and $\bar{m}(T)=m_0e^{-q_m/k_BT}$ \cite{Arrhenius1889}, where $D_0$ and $m_0$ are the temperature-independent parameters and $q_D$ and $q_m$ are activation energies,  respectively. These were obtained by fitting the values calculated via MD simulations to the above expressions, in the temperature interval of $300-1000$ K.}
\begin{center}
\begin{tabular}{|M{10cm}|M{4.0cm}|}
\hline
\textbf{Property} & \textbf{Value} \\
\hline
 Temperature-independent diffusion coefficient, $D_0$ & $3.80\cdot 10^{-3}$ cm$^{2}$ s$^{-1}$\\
 Activation energy of the diffusion coefficient, $q_D$ & $0.46$ eV\\
 Formation enthalpy, $h_v$ & $1.21$ eV \\
 Mixing term $f_2$ & $0.07$ eV \\
 Mixing term $f_3$ & $-3.47$ eV \\
 Mixing term $f_4$ & $2.19$ eV \\
 Grain boundary energy, $\sigma_0$ & $2410$ mJ m$^{-2}$ \\
 Temperature-independent grain boundary mobility, $m_0$ & $1.5$ m s$^{-1}$MPa$^{-1}$ \\
 Activation energy of the  grain boundary mobility, $q_m$ & $0.027$ eV\\
 Molar volume, $V_m$ & 41.03 cm$^3$ mol$^{-1}$ \\
 Sink strength at grain boundaries, $s^{GB}_v$ & $0.02$ ns$^{-1}$ \\
 Vacancy generation/annihilation rate, $s^{G}_v$ & $1$ ns$^{-1}$\\
\hline
\end{tabular}\end{center}
\end{table}

The objective of this section is to study grain growth and the mean grain size distribution in the presence of immobile voids fixed in size which block grain motion. Simulation boxes of different sizes and number of grains were used to confirm the independence on the box size and the final results were obtained from the statistical average over different simulations.  The evolution of the mean grain size, $\langle r(t) \rangle$, is described by the general equation \cite{Burke1952}
\begin{equation}
\frac{d\langle r(t) \rangle}{dt}=\frac{k}{n\langle r(t) \rangle^{n-1}},
\label{eq:paraboliclawder}
\end{equation}
where $k$ is the kinetic coefficient and depends on the material and temperature. The exponent $n$ depends on the material and is around $2$ in pure materials \cite{Burke1952,Humphry-Baker2014}. According to this equation, the mean grain size grows until the single crystal is reached:
\begin{equation}
\langle r(t) \rangle^n-\langle r(t_0) \rangle^n=k(t-t_0).
\label{eq:paraboliclaw}
\end{equation}
We analyzed grain growth at different void fractions, $d_v$, in  polycrystalline PbTe using phase-field simulations at 500 K and 300 K. We found that $n$ is $1.94$ in void-free polycrystalline PbTe ($d_v=0$), with $k$ following an Arrhenius law in temperature \cite{Arrhenius1889}. The grain size follows Hillert's distribution \cite{Hillert1965} in the steady-state regime. At bulk vacancy concentrations above the equilibrium concentration, $n$ remains constant and grain growth does not stop. 

\begin{figure}
\centering
        \includegraphics[width=0.75\textwidth]{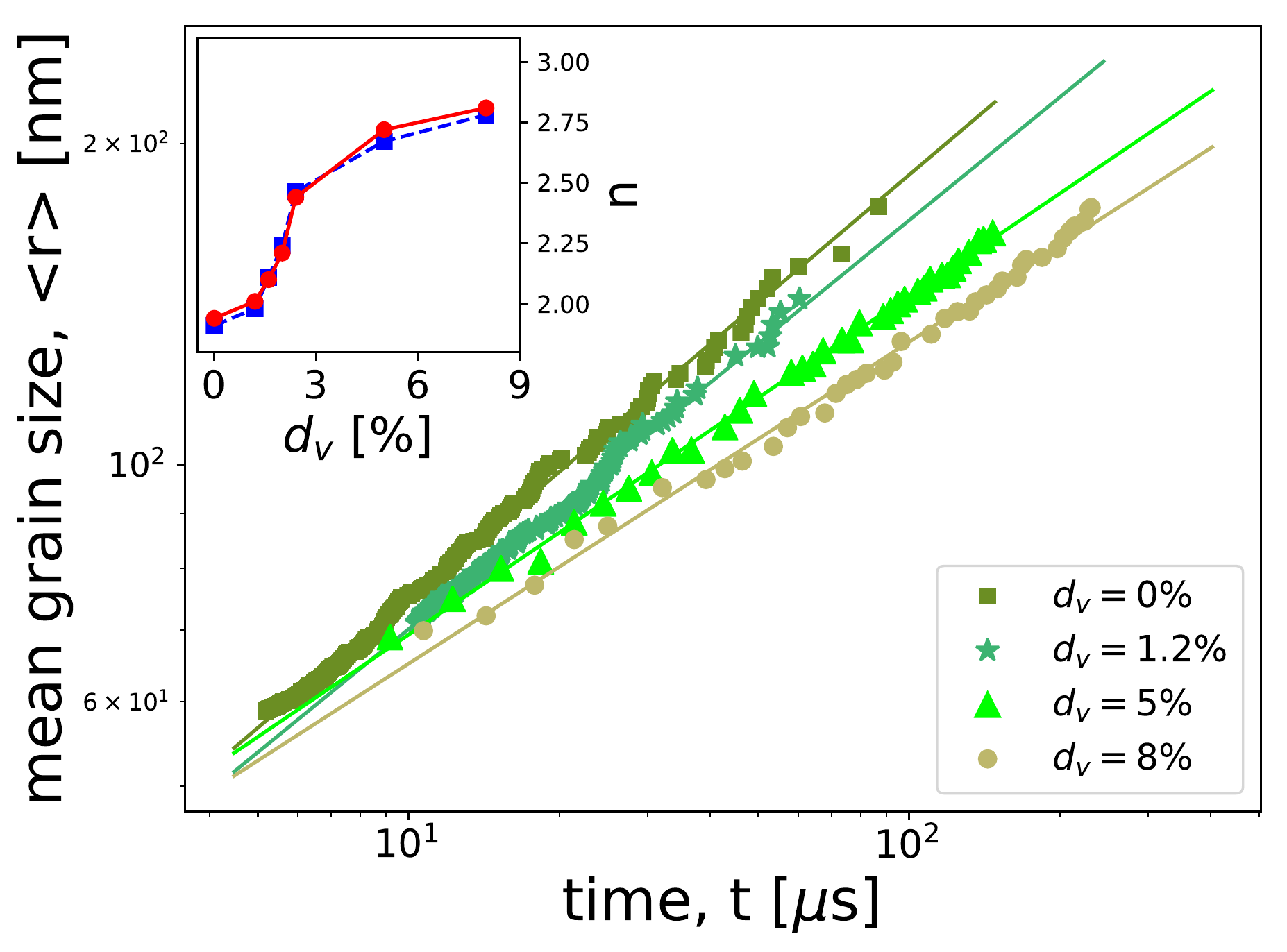}
        \caption{\small Evolution of the mean grain size over time at different void fractions, $d_v$, at $500$ K with constant void radius, $r_v=45$ nm, at the steady state regime where Eq.~\ref{eq:paraboliclaw} is valid. The void fraction is defined as the percentage of grid points occupied by voids and the vacancy concentration in bulk is $c_v=10^{-4}$. Lines correspond to Eq.~\ref{eq:paraboliclaw}. Inset: the exponent $n$ (Eq. \ref{eq:paraboliclaw}) depends on $d_v$ and shows a transition from the regime where voids are sufficiently separated ($\lessapprox 2\%$) to a regime where the spacing between voids becomes comparable to grain size ($\gtrapprox 5\%$). The dependence of $n$ on $d_v$ is the same at 300 (blue squares) and 500 K (red dots). 
        }
        \label{fig:expgrainwvoid300K}
\end{figure}

\begin{figure}[h!]
        \centering
        \includegraphics[width=0.7\textwidth]{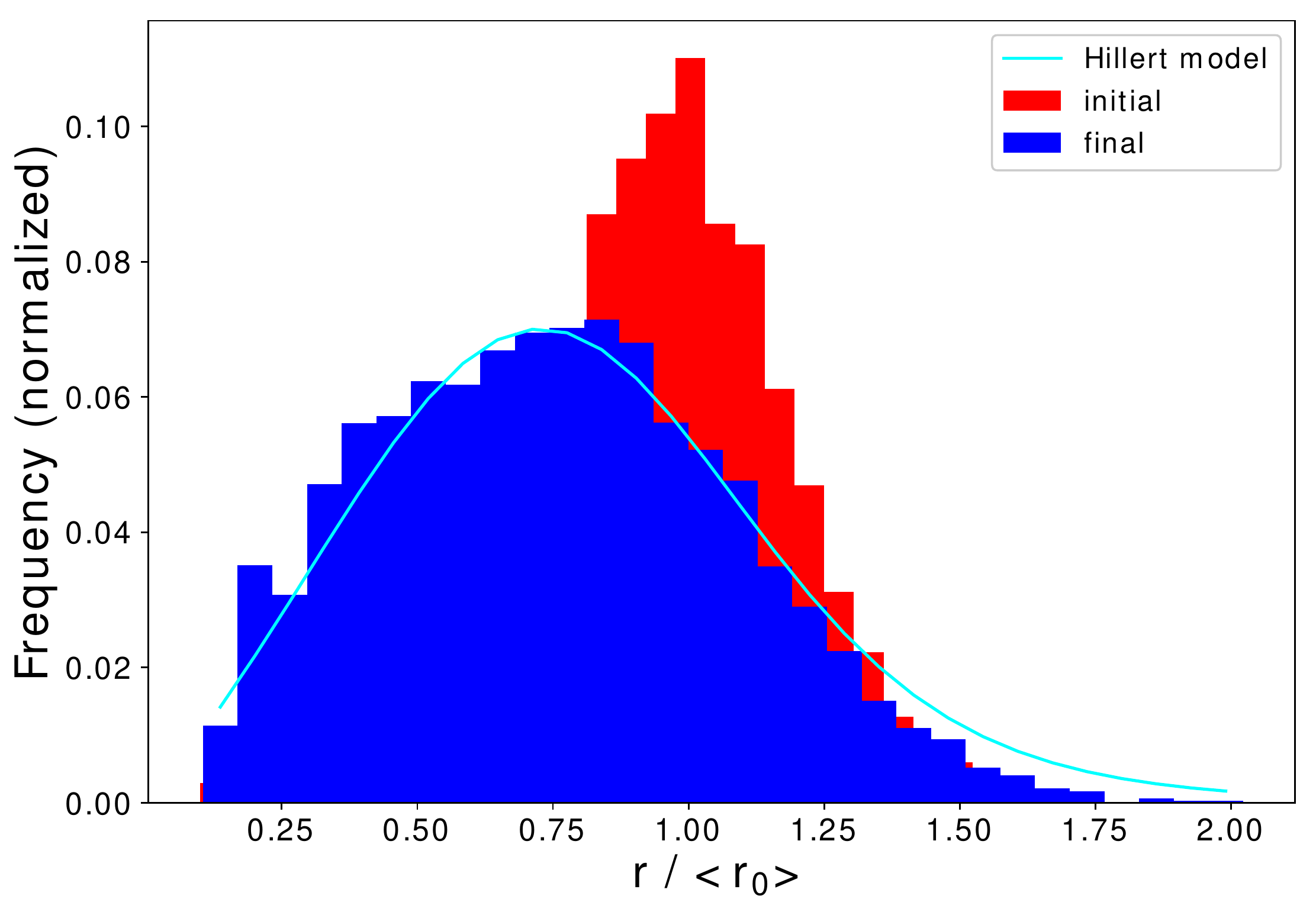}  
        \caption{\small Evolution of the grain size distribution at the initial configuration (red) the steady state regime (blue), when it can be described by the Hillert's distribution with $\bar\rho=1.0\pm 0.05$ and $\bar\gamma=2.12 \pm 0.03$. This simulation corresponds to the polycrystal under the equilibrium vacancy concentration at 500 K after 50 $\mu$s.}
        \label{fig:hillertdist}
\end{figure}

\begin{figure}[h!]
\centering
        \includegraphics[width=0.7\textwidth]{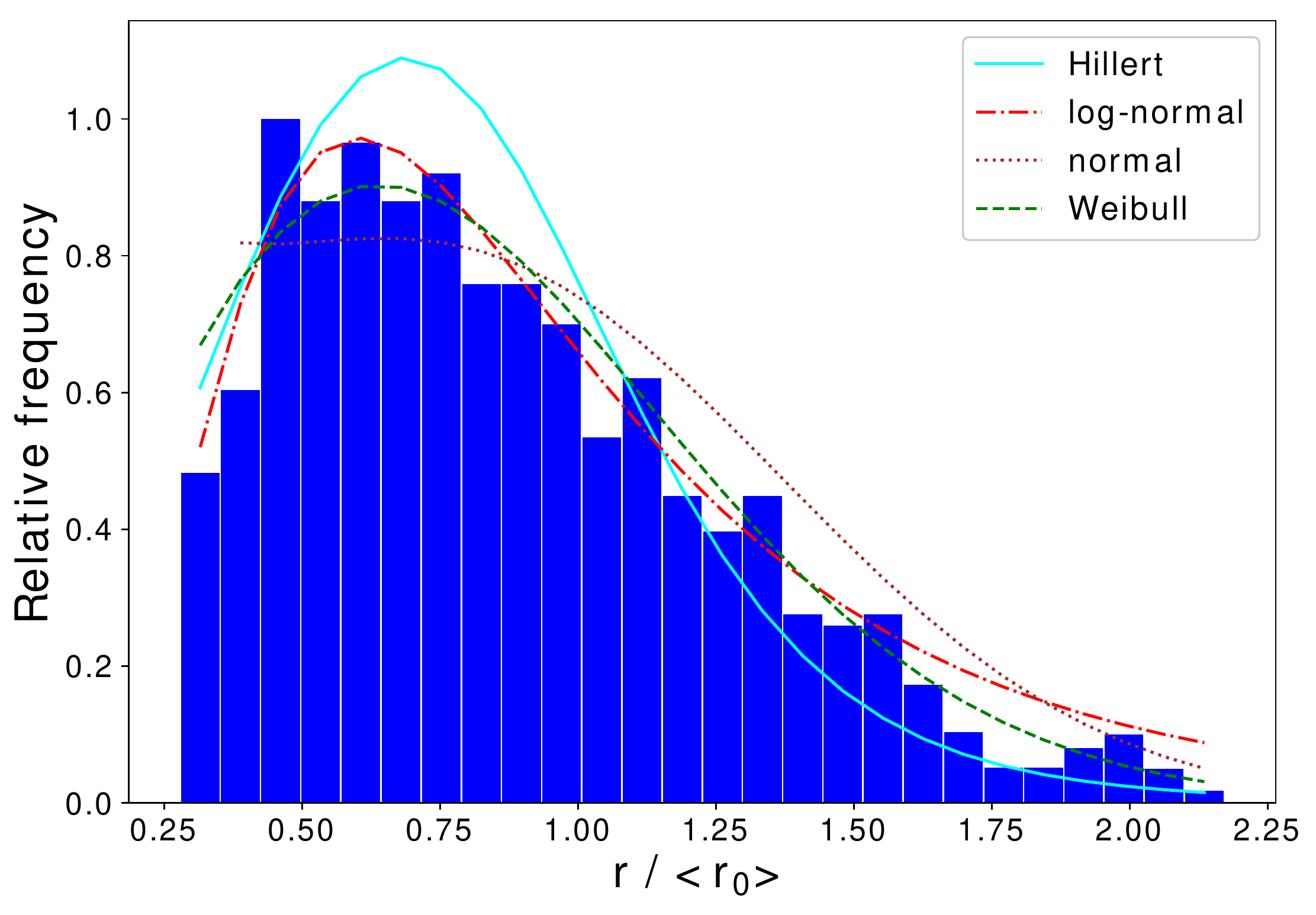} 
        \caption{\small Evolution of the grain size distribution in the steady state regime in the presence of voids ($d_v=8\%$) at 300 K. This simulation corresponds to the saturated polycrystal at 300 K after 100 $\mu$s. Different distribution functions found in literature are compared: Hillert's \cite{Hillert1965}, log-normal \cite{Okazaki1972}, normal and Weibull's distribution \cite{Korbuly2017}. The best match is obtained with the log-normal distribution. }
        \label{fig:graindistwvoids}
\end{figure}

We observed that the exponent $n$ in Eq.~\ref{eq:paraboliclaw} depends on void fraction, as can be seen in Fig.~\ref{fig:expgrainwvoid300K} for a fixed void size of $45$ nm. For void fractions $d_v=0-1.2\%$, grain growth does not differ from the void-free system and $n=1.94$ is the best fit in Eq.~\ref{eq:paraboliclaw}. However, $n$ increases with void fraction for $2\%<d_v<5\%$, asymptotically reaching $n\rightarrow 3$ for the largest $d_v$; we found $n=2.77$ for $d_v=8\%$. The same dependence is found at 300 K. Furthermore, as the exponent changes, the mean grain size distribution shifts from Hillert's distribution when $n\approx2$ to a log-normal distribution at $n\approx3$. Hillert's distribution is given by \cite{Hillert1965}:
\begin{large}\begin{eqnarray}
f\left(\frac{r}{\langle r_0 \rangle}\right)=\frac{3\bar\gamma^{3/2}\bar\rho\frac{r}{\langle r_0 \rangle}}{\left( \left(\bar\rho\frac{r}{\langle r_0 \rangle}\right)^2- \bar\gamma\bar\rho\frac{r}{\langle r_0 \rangle} +\bar\gamma\right)^{5/2}} \nonumber\\ 
\times e^{-\frac{3\sqrt{\bar\gamma}}{\sqrt{4-\bar\gamma}}\left(\arctan\left(\frac{2\bar\rho\frac{r}{\langle r_0 \rangle}-\bar\gamma}{\sqrt{\bar\gamma(4-\bar\gamma)}}\right)+\arctan\left(\frac{\bar\gamma}{\sqrt{\bar\gamma(4-\bar\gamma)}}\right)\right)},
\end{eqnarray}\end{large}
\noindent where $r$ is the grain size, $\langle r_0 \rangle$ is the mean grain size and $\bar\gamma$ and $\bar\rho$ are fitting parameters. This model describes normal grain growth, as we can see in Fig. \ref{fig:hillertdist} in the absence of voids ($d_v=0\%$). In Fig. \ref{fig:graindistwvoids}, we observe that the log-normal distribution is a better fit in the presence of voids. Dispersion and secondary peaks at the tail can be due to the presence of voids and anisotropic grain boundaries. This deviation from Hillert's distribution is most likely caused by the breaking of scale-invariance shown in Fig. \ref{fig:expgrainwvoid300K} (inset), on which Hillert's distribution relies.
Since Hillert's distribution was derived assuming that it converges to a self-similar fixed point, this departure signals a characteristic size in the higher $d_v$ regime. The power-law growth in time, observed upon scale-invariant phase-field simulations, can be associated with the universal exponent $\nu=1/n$. Therefore, Fig.~\ref{fig:expgrainwvoid300K} reveals a change in the critical exponent from $\nu=1/2$, characteristic of a Gaussian fixed point, towards an interacting theory with $\nu\approx 1/3$. This implies a finite void-void interaction in the high-$d_v$ fixed point. By Widom-Rushbrooke scaling relations we expect all other critical exponents to be modified and other measurable quantities, e.g. thermal conductivity, to be affected. Most importantly, a change in $n$ highlights that the kinetics of impeding grain growth by Zener pinning is a collective phenomenon governed by a phase transition. The existence of a tentative critical point suggests that this mechanism may be at play in more general situations, and sheds light into the question of why polycrystals are so abundant in nature \cite{Holm2010a}.


\subsection{Void stability}\label{sec:voids}
In the present section, the phase-field model introduced in Section \ref{sec:methods} is used to study the stability of voids in bulk PbTe. In a vacancy-saturated regime, vacancies can form vacancy clusters (unstable) or voids (stable). If the cluster is smaller than a critical size, then it shrinks and disappears, while it grows when its radius is larger than a critical size and becomes a void. This critical size depends on temperature and vacancy concentration. 

The free energy required to form a spherical void of radius $r_v$ is given by classical nucleation theory (CNT) \cite{Tan1997}:
\begin{equation}
\Delta G_V^{CNT}=4\pi r_v^2\sigma_v-\frac{4\pi r_v^3}{3\Omega}k_BT\textnormal{ ln}\left(\frac{C_V}{C_V^{eq}}\right),
\label{eq:voidfreeenergy}
\end{equation}
where $\sigma_v$ is the interfacial energy, $\Omega$ is the volume per formula unit, $C_V$ is the total vacancy concentration and $C_V^{eq}$ is the equilibrium value at temperature $T$. This total vacancy concentration includes both vacancies in bulk and vacancies forming voids. In contrast with $c_v$, $C_V$ corresponds to the total concentration over the whole system, whereas $c_v$ is the local concentration at one specific grid point. If the simulation box is large enough, the total vacancy concentration can be approximated by the vacancy concentration in bulk. The first term in Eq.  \ref{eq:voidfreeenergy} corresponds to the increase in energy due to the formation of the void/solid interface and the second term is related to the work performed to relocate vacancies inside a new void. Interactions between voids and vacancies with vacancy sources, dislocations and grain boundaries are not considered in CNT. The critical or cut-off void radius corresponds to $d(\Delta G_V^{CNT})/dr_v=0$, i.e.

\begin{equation}
r_v^{cr}=\frac{2\Omega\sigma_v}{k_BT\textnormal{ ln}\left(\frac{C_V}{C_V^{eq}}\right)}.
\label{eq:voidcrsize}
\end{equation}

The Cahn-Hilliard equation (Eq. \ref{eq:cahnhilliard}) was solved numerically using finite differences and the forward Euler time integration method to study the stability of voids in vacancy-saturated PbTe, i.e. in a system where the vacancy concentration in bulk is higher than its equilibrium value. A single void was placed in the center of the simulation box and we observed how its size evolved over time. This study was repeated using different initial void sizes. The critical radius $r_v^{cr}$ was determined by identifying the largest initial radius for which the voids dissolve \cite{Li2011}. Simulations were run at different temperatures ($300-900$ K) and under different bulk vacancy concentrations ($c_v^{bulk}=0.1$, $0.05$, $0.01$). We observed that voids are stable at intermediate temperatures but not at high temperatures because then the high vacancy mobility  promotes diffusion and can more easily destabilize voids, especially small ones. At low temperatures, below 300 K, the mobility is very low and vacancies take longer to accumulate. On the other hand, it is observed that if the number of vacancies is low, it is more difficult to create voids and their cut-off radius is larger.  

\begin{figure}[h!]
        \centering
        \includegraphics[width=0.7\textwidth]{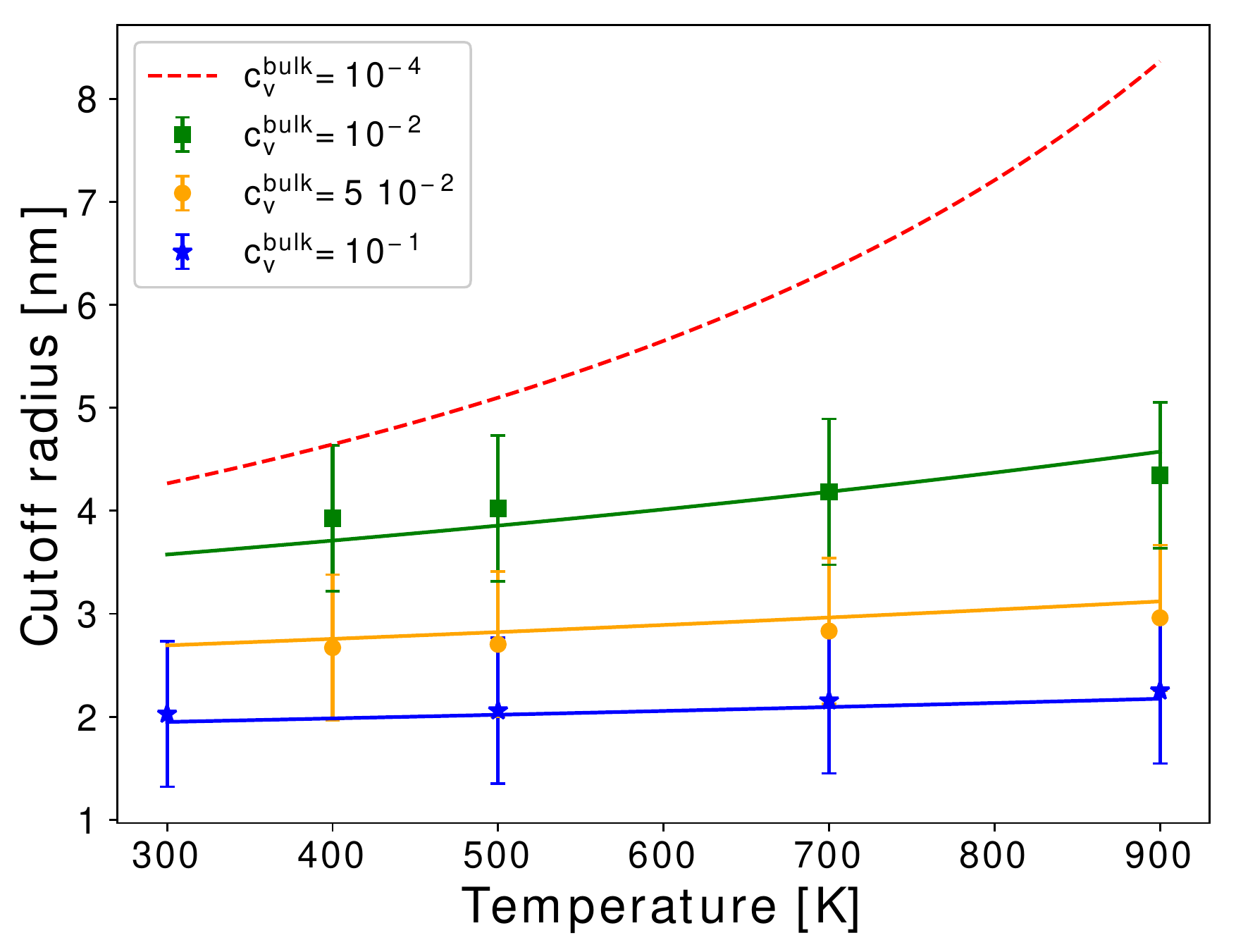} 
        \caption{\small Critical radius (void size) at different temperatures and different vacancy concentrations. Voids smaller than this cut-off size dissolve, while voids with a radius larger than the cut-off are stable and can grow. At higher temperatures, the vacancy diffusion coefficient is higher and vacancies annihilate faster. Hence, small voids are destabilized, and the cut-off radius is larger. On the other hand, at higher vacancy concentrations, smaller concentration gradients promote the existence of smaller voids. Lines correspond to Eq. \ref{eq:voidcrsize}. 
        }
        \label{fig:voidsize}
\end{figure}

These effects can be seen in Fig. \ref{fig:voidsize}, where we report the critical radii determined as explained above, as a function of concentration and temperature (symbols with error bars). Since the simulation box is large enough for the vacancy concentration in bulk to be roughly constant, we have fitted the simulation results with Eq. \ref{eq:voidcrsize} (lines in Fig. \ref{fig:voidsize}). The fitting parameters are $\sigma_v(T,C_V)$ and $C_V(T)$. The time required to equilibrate voids in the presence of small vacancy concentrations can be hours or days. Therefore, in order to access the experimentally realistic regime, we adopted the following strategy: we first calculated $r_v^{cr}(T)$ for a set of larger concentrations (0.1, 0.05, and 0.01), and then extrapolated to a bulk vacancy concentration $c_v^{bulk}=10^{-4}$ using Eq. \ref{eq:voidcrsize} (red curve in Fig. \ref{fig:voidsize}) using $\sigma_v$ and $C_V$ obtained at higher concentrations. At 500 K, the critical radius would be around $5$ nm.

Equation \ref{eq:voidfreeenergy} predicts either dissolution or nucleation of voids, but not metastable finite void sizes. However, multiple metastable finite-size voids are found in experiments \cite{Yoon2013}, and this expression needs to be modified to account for this. Assuming that deviations from CNT arise from the presence of multiple voids, we study a system containing two voids and introduce an additional phenomenological term in the free energy to account for the effect of interactions between voids
\begin{equation}
\Delta G_V=2\cdot 4\pi r_v^2\sigma_v-2\cdot \frac{4\pi r_v^3}{3\Omega}k_BT\textnormal{ ln}\left(\frac{C_V}{C_V^{eq}}\right)+\gamma_vr_v^s,
\label{eq:voidfreeenergyfinal}
\end{equation}
with $\gamma_v$ and $s$ constants. The first two terms are the same as in Eq.  \ref{eq:voidfreeenergy} but doubled as this expression corresponds to two voids. Terms like the last one in Eq. \ref{eq:voidfreeenergyfinal} have been proposed for other systems exhibiting nucleation of secondary particles where $s$ is in the range $3-5$ \cite{Gusak2011,Hodaj2004,Desre1990}.   If $\gamma_v$ is chosen such that a second minimum with $\Delta G_V=0$ emerges at finite $r_v$, then systems containing no voids and systems containing voids of this metastable size are equally favorable.

\begin{figure}[h!]
        \centering
        \includegraphics[width=0.6\textwidth]{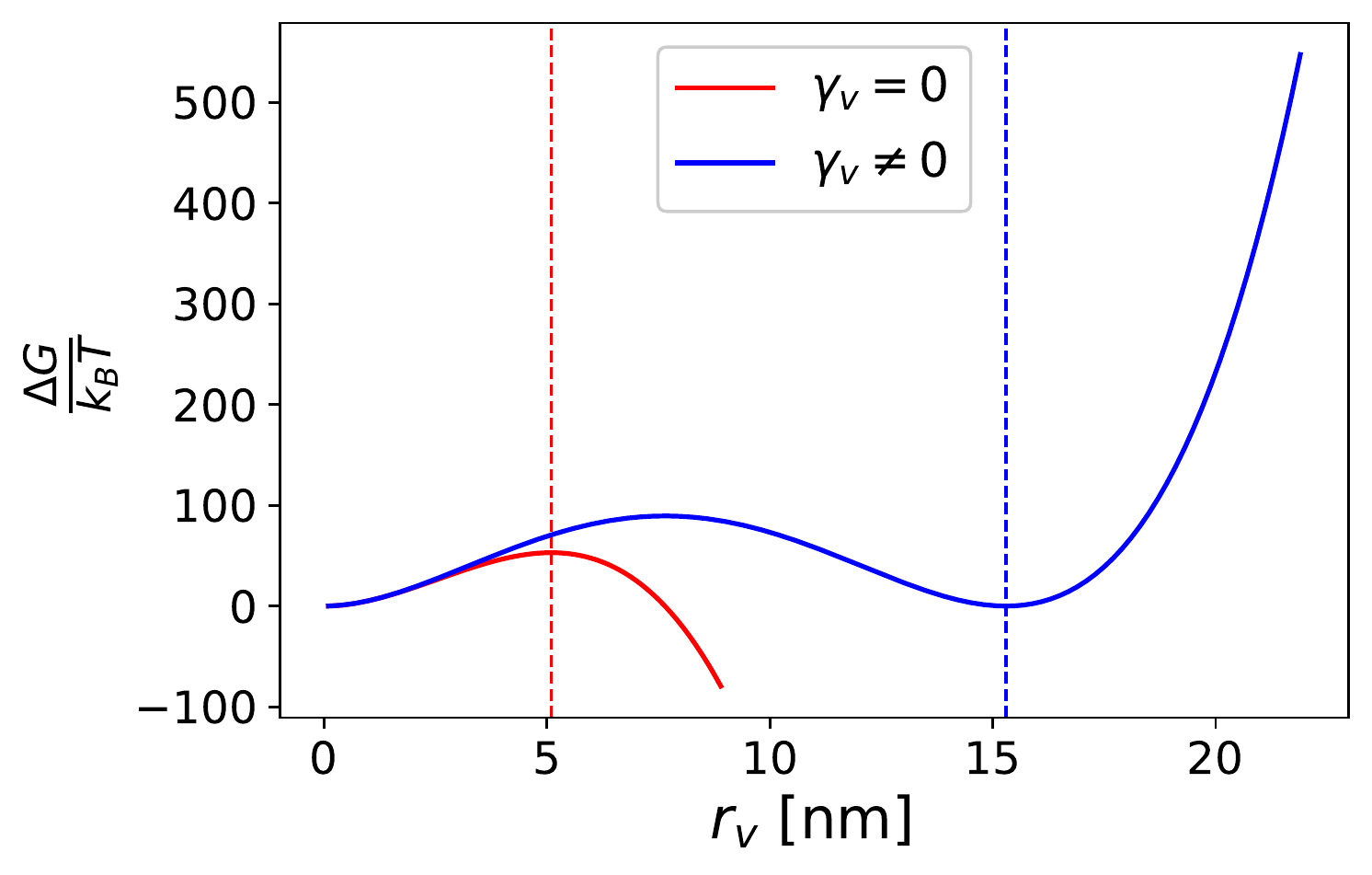} 
        \caption{\small Free energy of a system containing two voids of radius $r_v$ at 500 K. Comparison between the standard model given by Eq. \ref{eq:voidfreeenergy} (red), in which there is no metastable finite void size, and Eq. \ref{eq:voidfreeenergyfinal} (blue), where a metastable size appears at $15$ nm.
        }
        \label{fig:voidfreeenergyfinal}
\end{figure}

We used the phenomenological Eq. \ref{eq:voidfreeenergyfinal} to gain insight into the behavior of metastable voids in real applications. In a system with a bulk vacancy concentration of $c_v^{bulk}=10^{-4}$, the critical radius at 500 K is approximately 5 nm (see dashed red line in Fig. \ref{fig:voidsize}). This critical radius corresponds to a maximum in the free energy, as shown by the red line in Fig. \ref{fig:voidfreeenergyfinal}, which corresponds to non-interacting voids (Eq. \ref{eq:voidfreeenergyfinal}).  Keeping parameter $\sigma_v$ in Eq. \ref{eq:voidfreeenergyfinal} the same as in CNT (Eq. \ref{eq:voidfreeenergy}) and choosing a suitable value of $\gamma_v$, we can observe the emergence of a minimum in the free energy at a finite size, appearing at 15 nm in the blue line in Fig. 
\ref{fig:voidfreeenergyfinal}. This minimum represents a metastable void size, meaning that voids created with this size should neither dissolve nor nucleate. In addition, the critical size (location of the maximum in the blue line) increases from 5 to 8 nm due to the presence of multiple voids simultaneously, i.e. due to void-void interactions. Experimental studies confirm the presence of such metastable voids in real-life PbTe samples \cite{Yoon2013}.

\begin{figure}[h!]
\centering
\includegraphics[width=0.75\linewidth]{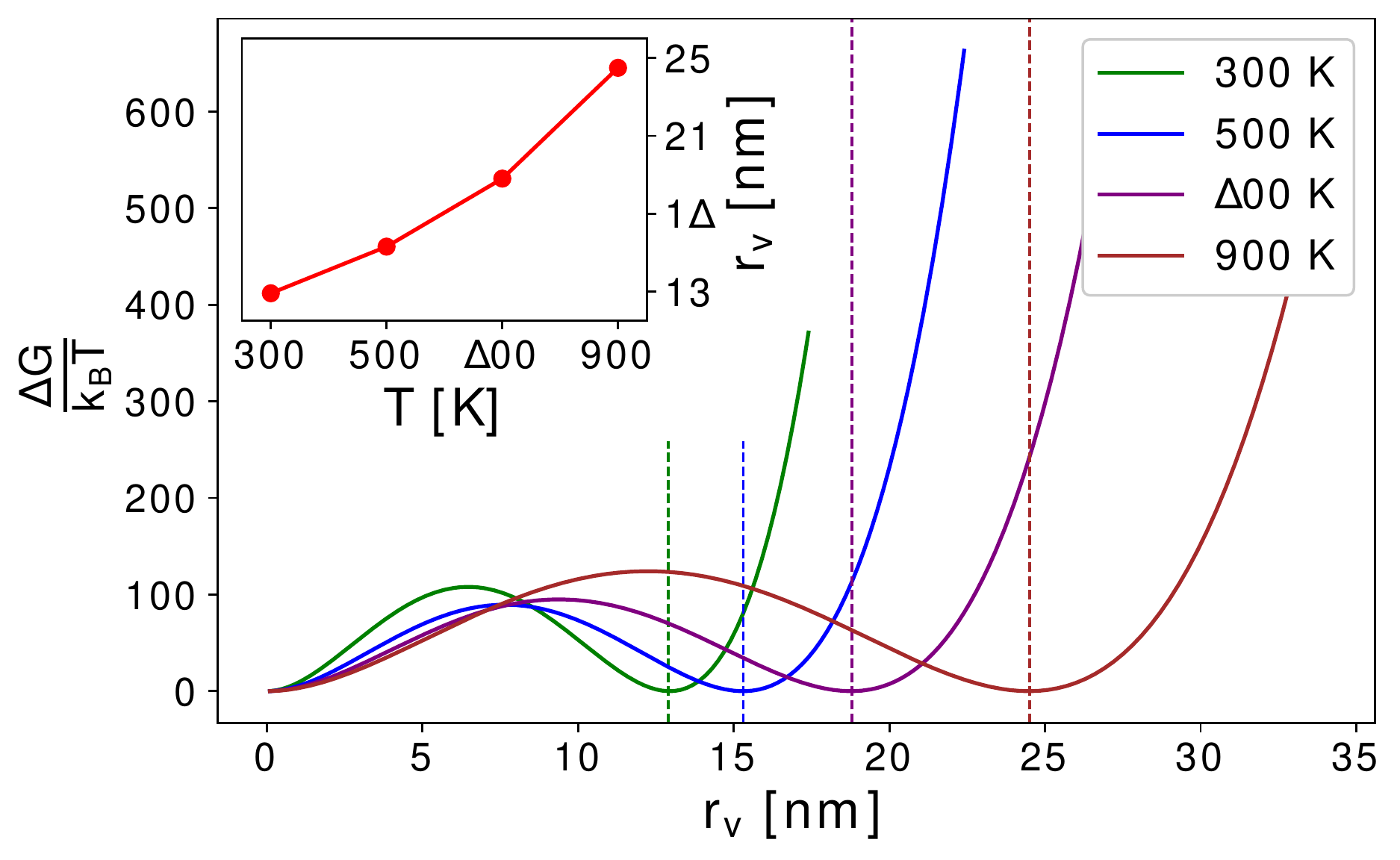}
        \caption{\small Free energy of a system containing two voids of radius $r_v$ at different temperatures according to Eq. \ref{eq:voidfreeenergyfinal}. Inset: evolution of the metastable void size with temperature. 
        }
\label{fig:metastablevoid}
\end{figure}

The dependence of the metastable size on temperature is shown in Fig. \ref{fig:metastablevoid}, for a bulk vacancy concentration of $10^{-4}$ and an interfacial energy $\sigma_v$, determined via Eq. \ref{eq:voidcrsize}, using the critical radius extracted from Fig. \ref{fig:voidsize}.  These metastable sizes, which increase with temperature as shown in the inset to Fig. \ref{fig:metastablevoid}, correspond to the formation of voids when the metastable size is forced to be at $\Delta G_V=0$. Metastable sizes with $\Delta G_V\neq0$ may also be feasible, although these sizes would not be as energetically favorable as the void-free system.

\subsection{Zener pinning by voids}\label{ssec:c5_zenerpinning}

In Section \ref{sec:mgs} voids were considered as immobile second-phase particles whose size was fixed. The results in Fig. \ref{fig:expgrainwvoid300K} were obtained in the presence of voids with a fixed radius, $r_v=45$ nm. In contrast, in Section \ref{sec:voids} we studied void stability in the absence of grain boundaries, and showed, using a phenomenological model, that metastable voids of finite dimensions can be stabilized due to void-void interactions.
Inspired by this model we performed phase-field simulations of grain growth as in Section \ref{sec:mgs}, but now in the presence of voids of varying size. The rationale is that voids tend to act as effective pinning particles if the local radius of curvature of the grain boundary is relatively large and the void fraction is high enough \cite{Humphreys2004}. If $p_Z$ is the {\it reduced} pinning pressure, in units of inverse distance, exerted by Zener particles in a saturated system, grain growth in the presence of voids can be modeled by Eq. \ref{eq:paraboliclawder} with $n=2$ \cite{Burke1952,Humphry-Baker2014}, which corresponds to the ideal case, and the effect of voids is incorporated through the pinning pressure as  \cite{Humphreys2004}:
\begin{equation}
\frac{dr}{dt}=k\left(\frac{1}{2r}-p_Z\right),
\label{eq:paraboliclawzener}
\end{equation}
where the pinning pressure depends on the void fraction, $d_v$, and void radius, $r_v$, as \cite{Humphreys2004}:
\begin{equation}
p_Z=\frac{d_v^{a(d_v)}}{\alpha r_v},
\label{eq:redpressure}
\end{equation}
with $r_v$ the void radius and $\alpha$ a constant that accounts for deviations observed in experiments. Values of $\alpha$ between $2.7$ and $3.6$ were reported in experiments in alloys \cite{Humphreys2004} at high void fraction. The exponent $a(d_v)$ changes from $a(d_v)=1$ at low void fraction, where voids act as independent scattering centers, to $a(d_v)=1/3$ at high $d_v$, where voids act collectively. While Eq. \ref{eq:paraboliclawder} is a good approximation for small deviations from the ideal grain growth through the parameter $n$, Eq. \ref{eq:paraboliclawzener} studies grain growth when it is slowed down in the presence of voids until a limiting grain radius, $R_Z$, is reached and then it stops when the RHS vanishes. Using $p_Z$ as defined in Eq. \ref{eq:redpressure}, Eq. \ref{eq:paraboliclawzener} leads to an equilibrium grain radius $R_Z$ in the saturation limit of
\begin{equation}
R_Z=\frac{\alpha r_v}{2d_v^{a(d_v)}}.
\label{eq:limitinggrainsize}
\end{equation}

According to Eq. \ref{eq:limitinggrainsize}, the stabilization of small mean grain sizes is associated with the presence of small voids, i.e. to this end it is better to have many small voids than a few large voids. The study of large, slowly-growing grains carries a large associated computational cost. We therefore focused our research towards the study of grain growth in the presence of relatively small void radii and high void fractions. In addition, very small grains cannot be studied due to the finite resolution of the simulation grid ($\Delta x=0.643$ nm). In Section \ref{sec:mgs}, the void size studied ($r_v=45$ nm) was relatively large, corresponding to a limiting grain size $R_Z\approx 260$ nm. The grain sizes observed in these simulations, shown in Fig. \ref{fig:expgrainwvoid300K}, did not reach this size due to time limitations and the size of the simulation box. The fact is that, under these conditions, the second term in Eq. \ref{eq:paraboliclawzener} is much smaller than the first one, and hence the dynamics is dominated by an ideal diffusive behavior. In Eq. \ref{eq:paraboliclawder} in Section \ref{sec:mgs} we assumed that the Zener pressure was negligible. However, this approximation is not valid when the grain size is close to its limiting value. There, $p_Z$ is relatively large and cannot be ignored. In Section \ref{sec:mgs} we remedied the absence of the $p_Z$ term by allowing for deviations from the ideal case ($n=2)$ through the variation of the exponent $n$ in Eq. \ref{eq:paraboliclawder}, finding that $n$ increases with the void fraction. 

\begin{figure}[b!]
\centering
        \centering
        \includegraphics[width=1\textwidth]{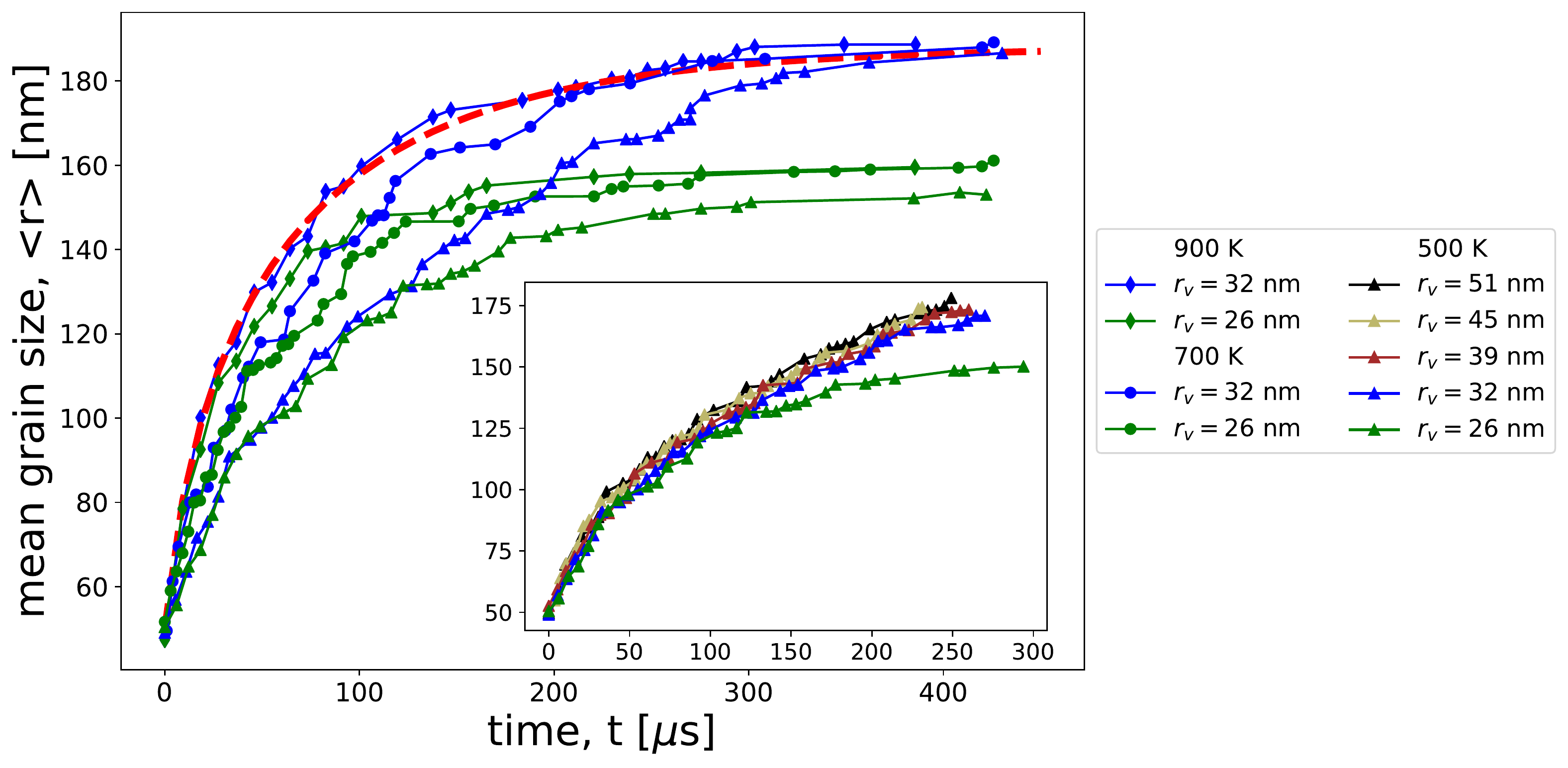} 
        \caption{Evolution of the mean grain size over time at constant void fraction, $d_v=8$ \%, at different void radii and temperatures in the steady-state regime. The vacancy concentration in bulk is $c_v=10^{-4}$. Dashed red line corresponds to Eq. \ref{eq:paraboliclawzener} for the data plotted using blue diamonds,  where $k$ and $p_Z$ are fitting parameters. The grain size would keep growing for hours until reaching an equilibrium value at low temperatures. In the inset, we can see the evolution of the mean grain size over time for different void sizes at 500 K. Due to the limitations in size and time associated with our simulations, the limiting grain size is not reached for the largest void sizes.
        }
        \label{fig:limitinggrainsize}
\end{figure}

In this Section we used Eq. \ref{eq:paraboliclawzener} instead, which keeps $n=2$ but includes the Zener pressure. We used smaller void sizes at high void fractions to be able to observe the grain growth slowing down, eventually reaching the limiting grain size. The time scale for observing this limiting behavior depends strongly on temperature. At low temperatures it can take hours to attain, and hence it is out of reach for the present simulations. For this reason we have focused on higher temperatures. In Fig. \ref{fig:limitinggrainsize} we show the time evolution of the mean grain size at 900, 700, and 500 K for the two smallest void sizes. The inset shows this evolution for a wider range of sizes, between 26 and 51 nm, at 500 K. Notice that, at 500 K, only for the smallest void size of 26 nm the simulations show the grains stopping growing in the time scale of the simulations (0.3 ms). According to Fig. \ref{fig:limitinggrainsize}, limiting sizes of $R_Z\approx180$ and $R_Z\approx155$ nm are obtained for void radii $r_v=32$ and $r_v=26$ nm, respectively. Comparing with Eq. \ref{eq:limitinggrainsize}, we obtain a value of $\alpha\approx 5$ at high void fraction, which is comparable to the values observed experimentally ($2.7-3.6$) \cite{Humphreys2004}. The difference may arise because these equations assume that all pinning particles interact with grain boundaries in the same way, but this is not the case. Voids can reduce the free energy much more than interstitial clusters formed by different chemical species in alloys and their impact on the elastic free energy is considerable. Although voids are known to be efficient pinning particles, the applicability of this equation has not been validated in their presence. Moreover, the materials used in experiments are not pure and can contain several pinning particles at the same time. 

\begin{figure}[b!]
        \centering
        
\subfigure[]{%
        \includegraphics[width=0.46\textwidth]{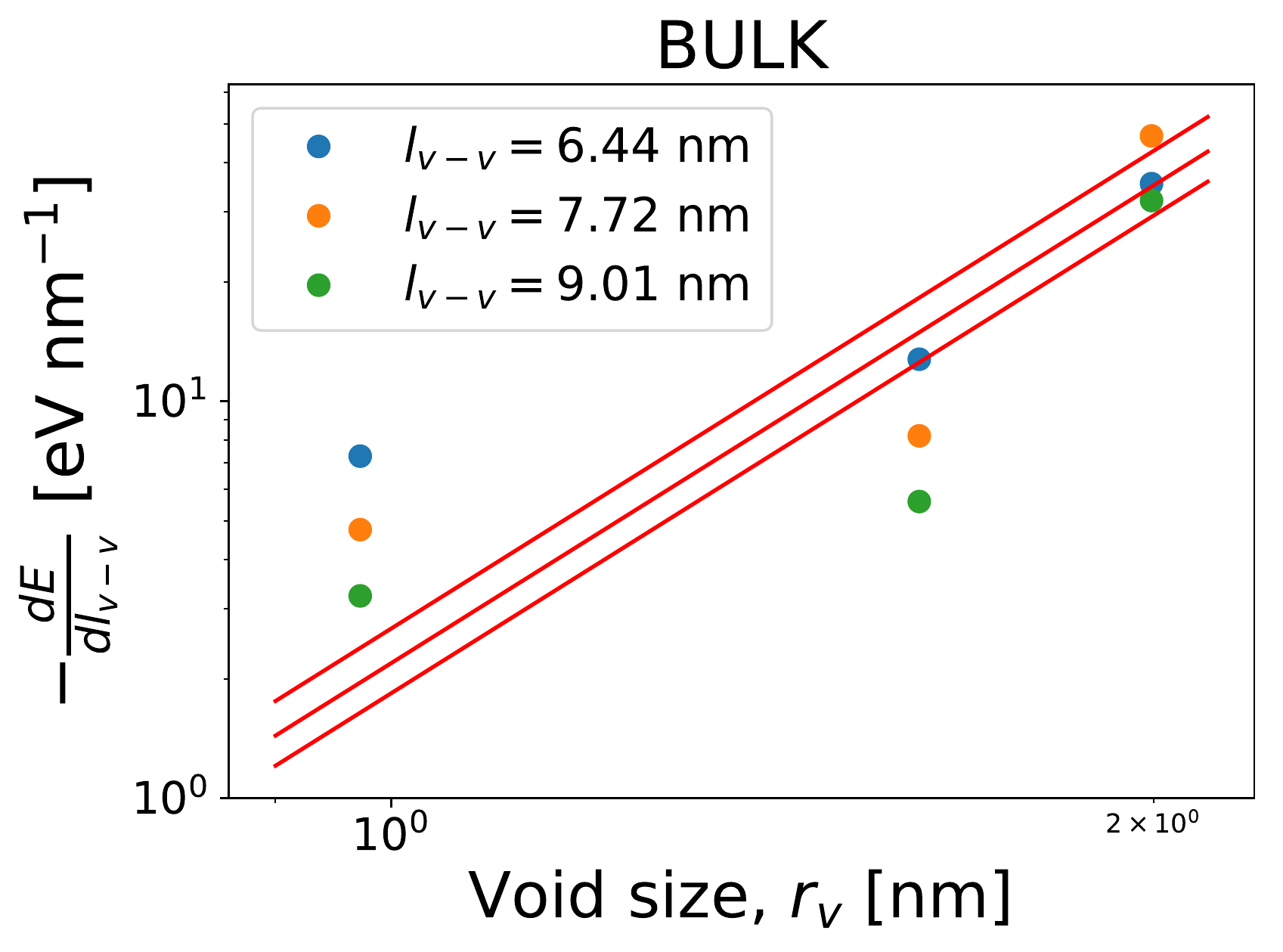} 
        }
\quad
\subfigure[]{%
        \includegraphics[width=0.46\textwidth]{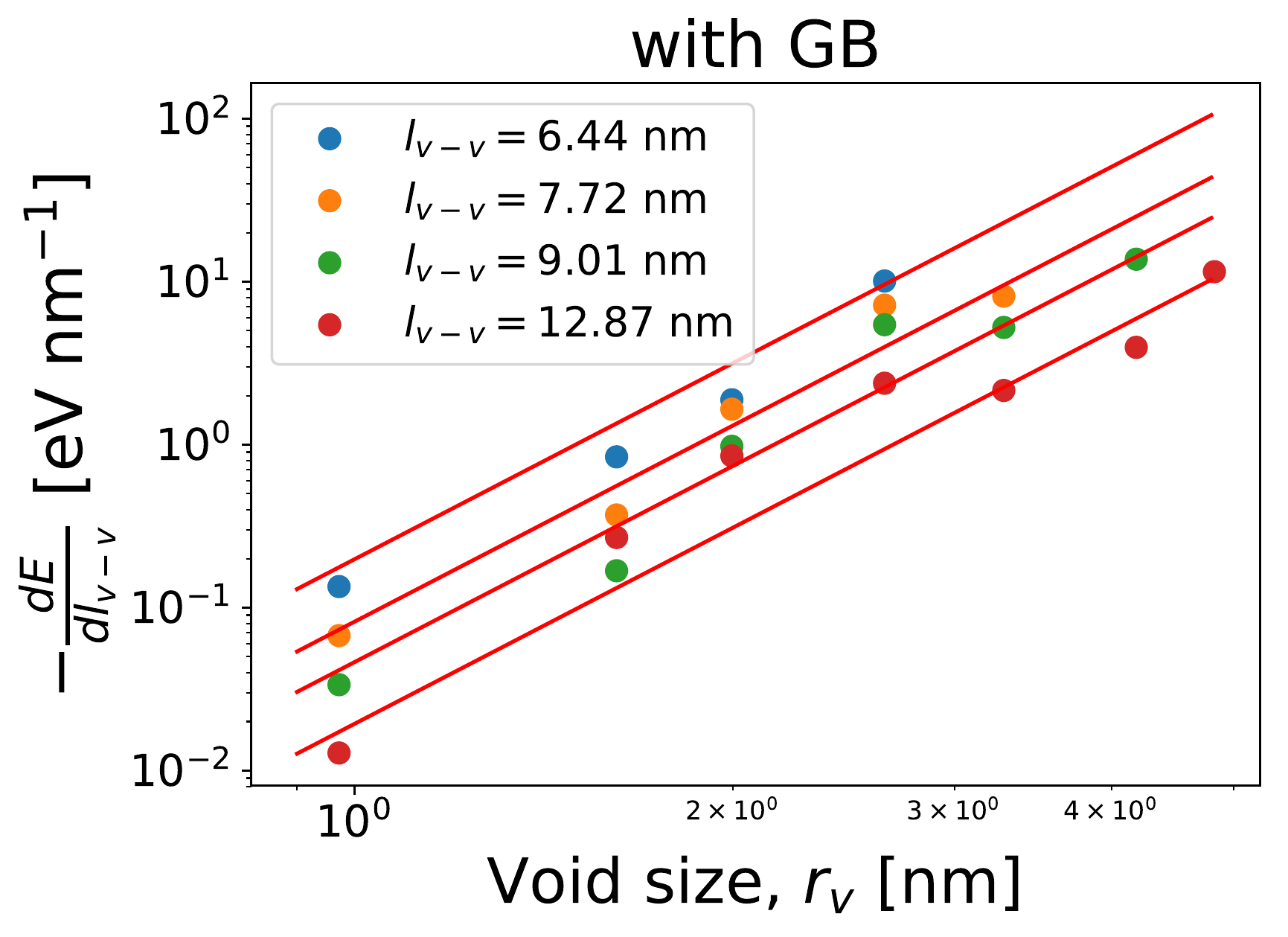} 
        }
        \caption{\small Analysis of the void-void interaction energy  in (a) bulk and (b) polycrystalline PbTe for different void sizes, $r_v$, and inter-void distances, $l_{v_v}$, calculated using LAMMPS via energy minimization. Two voids with equal size, $r_v$, are placed in the simulation box separated by a distance $l_{v_v}$. In the polycrystalline material, both voids are at the grain boundary. Red lines correspond to the functional form $-dE/dl_{v_v}=ar_v^4$, where $a$ is a fitting parameter. The force on each void associated with Zener pinning scales as $\propto r_v^4$ (b), while this dependence is not observed in the absence of grain boundaries (a). This implies the existence of a void-void interaction mediated by the grain boundary. 
        }
        \label{fig:voidinteraction}
\end{figure}

The presence of voids in the polycrystalline material results in an inter-void interaction mediated by grain boundaries, which can have a stronger effect in the presence of higher void densities. The force on each void associated with Zener pinning scales as $F_Z\sim p_Z r_v^2\sim r_v$ , where we used Eq. \ref{eq:redpressure}.  If we assume that the  void oscillation is related to phonons with characteristic energy $\omega_0$, the strength of the coupling for each void is given by $A_V = F_Z x_c / \omega_0$, where $x_c$ is  the characteristic interaction distance. The energy of an elastic domain wall (DW) scales like $R_Z^2$ \cite{Spaldin2010}, and from Eq. \ref{eq:limitinggrainsize} we see that $R_Z  \sim r_v$. Therefore, the indirect interaction energy between voids via grain boundaries is given by the energy of the elastic domain wall multiplied by the strength of the coupling squared, hence scaling as $E_{v-v}\sim A_V^2 R_Z^2 \sim r_v^4$,  where we used Eq. \ref{eq:limitinggrainsize}. This dependence was validated in Fig. \ref{fig:voidinteraction} (right panel), where we show the attractive force between voids mediated by the grain boundary, $-dE/dl_{v-v}$, where $l_{v-v}$ is the distance between two voids pinned at the grain boundary. The observation that the force is approximately linear with the logarithm of the void radius indicates a power-law behavior with $r_v$. The red lines in Fig. \ref{fig:voidinteraction} correspond to a fit with $r_v^4$. Apart from some small deviations, the good quality of this fit supports the argument that void-void interactions are mediated by grain boundaries and introduce a term of the form $\gamma r_v^4$ in the phenomenological expression for the free energy given by Eq. \ref{eq:voidfreeenergyfinal}. For comparison, we also report this force between voids in the absence of grain boundaries (left panel). It it clear that that this power-law scaling is not valid in this case. The inter-void interaction mediated by grain boundaries, in addition to the grain boundary-void interaction, are likely to be responsible for grain growth stagnation.

\subsection{Lattice thermal conductivity of porous polycrystalline PbTe}\label{ssec:c5_thermalcondfinal}
As a thermoelectric material, the efficiency of PbTe can be enhanced by reducing the lattice thermal conductivity. Voids and grain boundaries reduce the thermal conductivity by themselves, but small grains have to be stabilized to stop grain growth. We observed that this grain growth could be stopped in the presence of voids at grain boundaries, so now we are going to quantify the reduction of the thermal conductivity in the presence of metastable voids and grain boundaries. To obtain a solvable model, we assume that the majority of voids are at grain boundaries and $r_v$ and $d_v$ retain their values at $T_{CR}$. $T_{CR}$ is the coarsening temperature, defined as the maximum temperature to which the sample has ever been exposed. From our phase-field model, we obtain not only the mean grain size $\langle r (t) \rangle$ or diameter $\langle d (t) \rangle$, but also the distribution of grain sizes,  $\omega\left({d_i}/\langle d\rangle; T_{CR}\right)$, which, according to the results of Section \ref{sec:mgs}, follows approximately a log-normal law, where $d$ is the grain diameter. This enables us to compute the effective lattice thermal conductivity, $\kappa_\mathrm{eff}$, of porous polycrystalline PbTe. Assuming phonon confinement and averaging over grain sizes, the effective thermal conductivity of the ensemble is given by the following summation over all grain sizes present in the sample:
\begin{equation}
{\kappa_\mathrm{eff}(T,\chi_v; T_{CR})^{-1}}=
\sum_i\omega\left(\frac{d_i}{\langle d\rangle}; T_{CR}\right)\left[\frac{d_i}{d_i+\delta_{gb}(T)}\frac{1}{\kappa_\mathrm{b}(T)}+\frac{R_\mathrm{K}(\chi_v)}{d_i+\delta(T)}\right],
\label{eq:thermcond}
\end{equation}
where $\kappa_\mathrm{b}(T)$ is the thermal conductivity of the bulk material, $d$ is the grain diameter, $\langle d \rangle (T_{CR})$ is the mean grain diameter, $\delta$ is the grain boundary width, $R_\mathrm{K}$ is the Kapitza resistance, $\omega$ is the log-normal weight for grain size $d_i$ and $T$ is the working temperature. $\langle d \rangle (T_{CR})$ has memory as it depends on the thermal history of the sample. Given the metastable void size, $r_v$, from Fig. \ref{fig:metastablevoid} at temperature $T_{CR}$, the limiting grain diameter can be calculated by Eq. \ref{eq:limitinggrainsize}, $d=2R_Z$. Thus, the mean grain size of the sample is determined by $T_{CR}$, the maximum temperature to which the sample has ever been subjected. This $T_{CR}$ is not necessarily equal to the working temperature in a thermoelectric device, $T$. Note that while $\kappa_b(T)$ depends on the fast phonon dynamics, the processes of grain growth happen on much longer time scales described by our phase-field simulations and are irreversible.  

The grain boundary width $\delta_{gb}$ and Kapitza resistance $R_\mathrm{K}$ were studied using MD by the direct method with voids at the grain boundary \cite{Schelling2002,JavPabJor2019}. $\delta_{gb}$ increases with temperature as $\delta_{gb}=\delta_0(T_m-T)^{-1/2}$, where $T_m=924^\circ $C is the melting temperature \cite{Lide1998} and $\delta_0=289$ nm K$^{1/2}$ is a fitting parameter for data collected from Ref. \cite{JavPabJor2019}. $R_\mathrm{K}$ is inversely proportional to the heat capacity and is constant above the Debye temperature \cite{Yang2002}. MD simulations show that $R_\mathrm{K}$ in the presence of voids at the grain boundary depends on void coverage, $\chi_V$, as $R_\mathrm{K} = R_\mathrm{K}^{_\mathrm{nv}}(1-\chi_V)^{-1}$ where $R_\mathrm{K}^{_\mathrm{nv}}$ refers to the void-free case.

\begin{figure}[b!]
\centering
\includegraphics[width=0.9\linewidth]{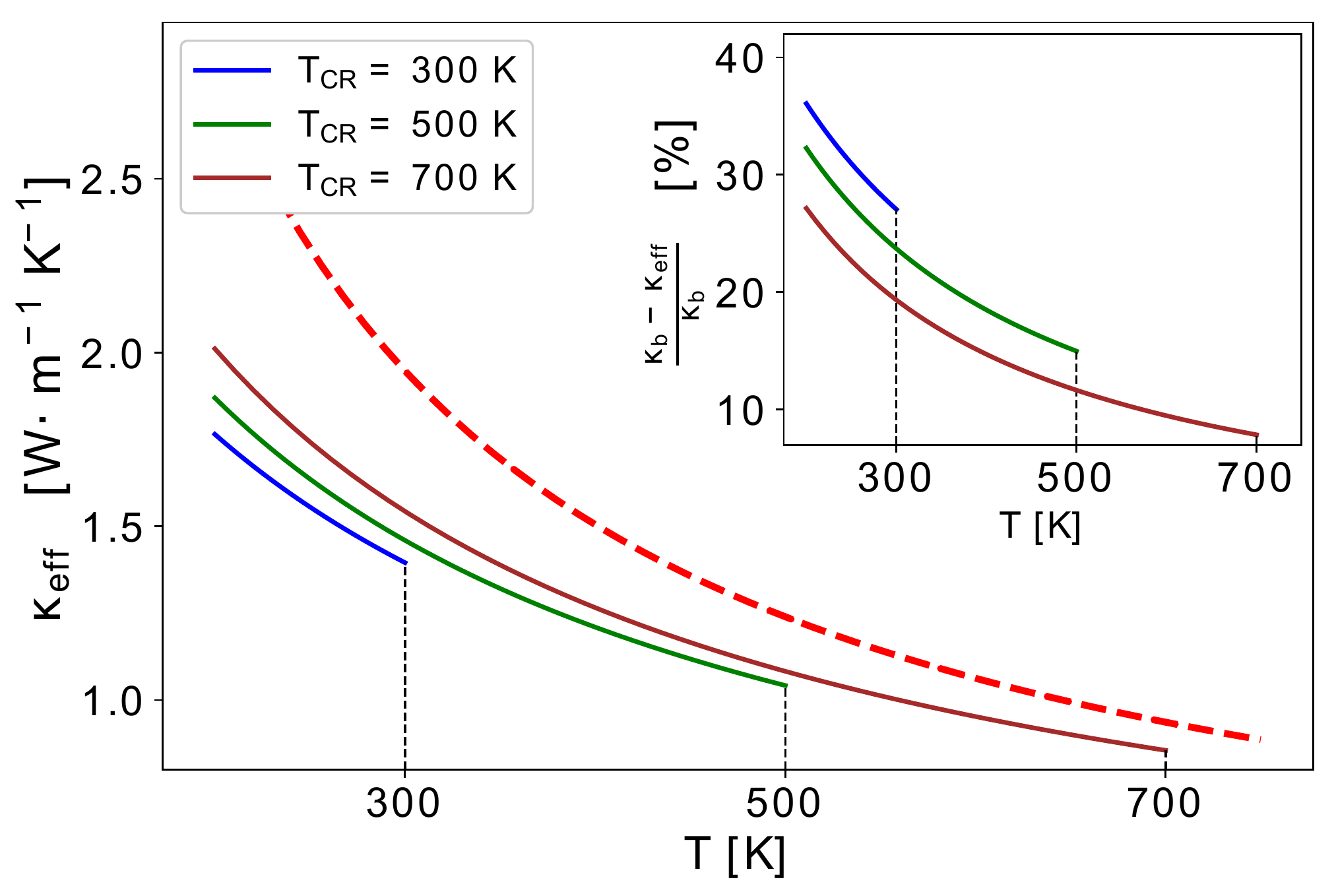}
        \caption{\small Effective thermal conductivity, $\kappa_\mathrm{eff}$, of porous polycrystalline PbTe as a function of temperature. Different curves are at different coarse-graining temperatures, $T_{CR}$, where $T_{CR}$ is defined as the maximum temperature to which the sample/device has been exposed. For the highest $T_{CR}$, we approach the limit of the bulk $\kappa_b(T)$ (dashed line), so comparing the curves allows assessing immediately the importance of the grain-size growth saturation at lower temperatures. The void fraction is $d_v=8\%$. Inset shows the relative loss of conductivity due to nanostructuring. 
        }
\label{fig:finalcondreduction}
\end{figure}

\begin{figure}[h!]
\centering
\includegraphics[width=0.8\linewidth]{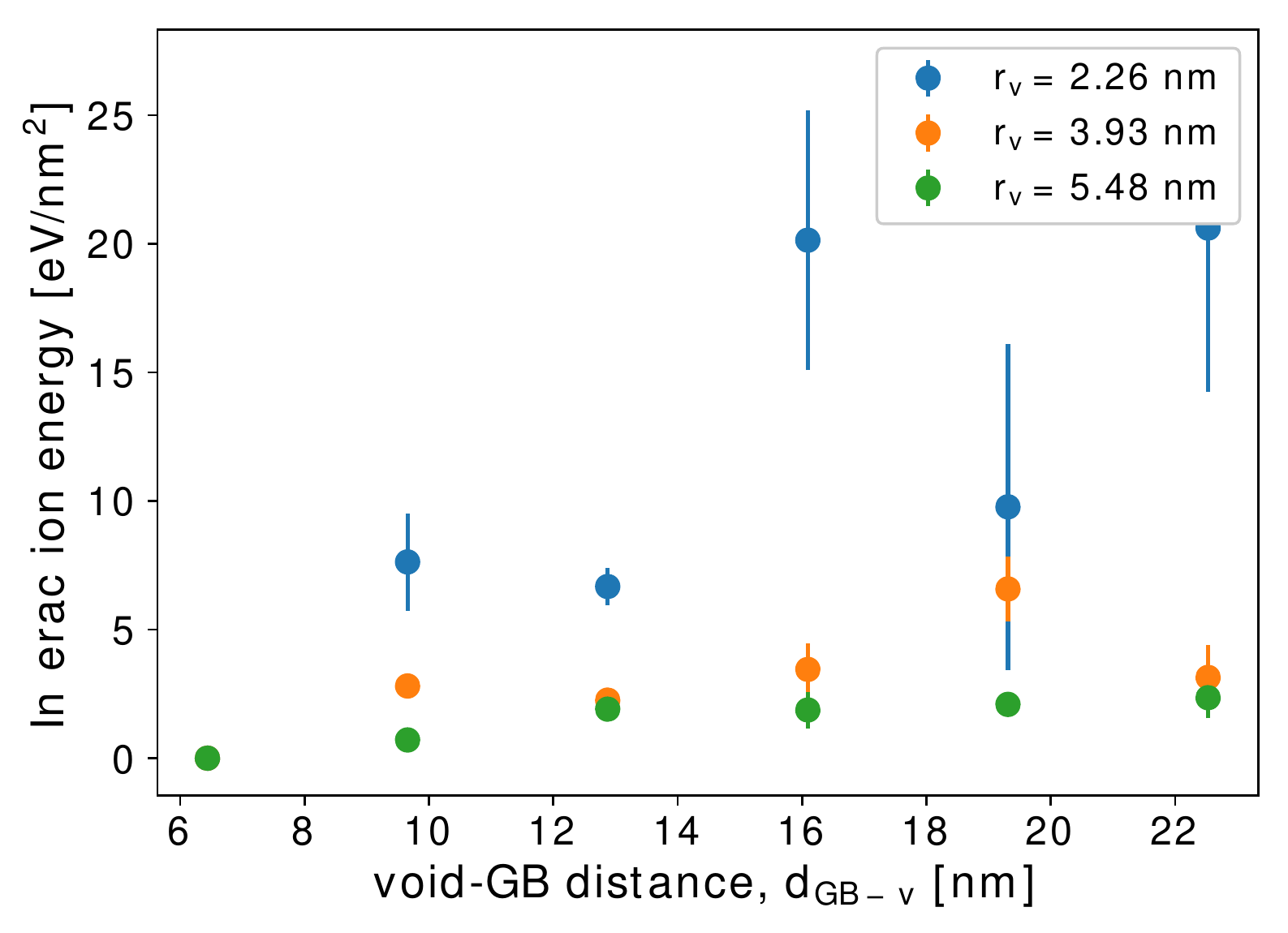}
        \caption{\small Void-GB interaction energy as a function of the void-GB distance, where GB stands for grain boundary and $r_v$ is the void radius. The interaction energy is defined as the excess energy divided by the void cross section, $\pi r_v^2$, when the void is far from the grain boundary and was calculated using LAMMPS after energy minimization. Systems in which voids are at the grain boundary are lower in energy and hence more stable.
        }
\label{fig:voidGBinteraction}
\end{figure}

We plot the result of Eq. \ref{eq:thermcond} in Fig. \ref{fig:finalcondreduction} at $d_v=8\%$. We show a family of curves $\kappa_\mathrm{eff}(T)$ for different thermal coarsening temperatures $T_{CR}$. This provides information about the thermal conductivity necessary to design thermoelectric devices with the only assumption that we work in thermodynamic equilibrium, where there is a well-defined phonon temperature $T$ at each point of the sample. We consider a constant working temperature along the length of the sample and $T_{CR}\geq T$ since the variation of the last quantity may be optimised separately.  A comparison between curves shows the effect of nanostructuring. The lattice thermal conductivity can be reduced nearly in half by grain boundaries pinned by voids. $\kappa_\mathrm{eff}(T)$ varies slower than in bulk, so the differences are largest at the lowest temperatures. Furthermore, our study shows the potential of introducing a coarse-graining profile $T_{CR}(x)$ inside the sample, going from blue to green to brown curves in different zones of the sample. This should reduce $\kappa_\mathrm{eff}(T)$ even more, enabling the design of some desired temperature profile. The regime at low void densities has not been studied here, but the validity of the existence of a limiting grain size at low densities of voids can be inferred from Eq. \ref{eq:limitinggrainsize}. Equation \ref{eq:thermcond} was evaluated assuming that voids are at the grain boundary, in which case their effect on the thermal conductivity enters through a change in the Kapitza resistance. This assumption is valid when the void fraction is high and the grain size is small. Additionally, energy calculations using the classical force field proposed in Ref. \cite{JavPabJor2019} were performed to validate this assumption. Results are reported in Fig. \ref{fig:voidGBinteraction}. It is clear that the energy is larger when voids are located farther away from the grain boundary (large $d_{GB-v}$), implying that configurations where voids are at the grain boundary (small $d_{GB-v}$) are more favorable, thus supporting the above assumption.

\section{Discussion}\label{sec:dicussion} 
Control over growth conditions is crucial to develop nanostructures for industrial purposes. PbTe is a semiconductor often used as a thermoelectric device, and its efficiency increases with reduced lattice thermal conductivities. As seen in Ref. \cite{JavPabJor2019}, the effective lattice thermal conductivity of PbTe falls in the presence of vacancies and grain boundaries, so special attention is given to these intrinsic defects. The vacancy concentration of single vacancies in PbTe can be estimated from electrical transport methods \cite{Schenk1988} and optical measurements \cite{Zhang2014}, and vacancy concentrations of up to $10^{-3}$ are found to be stable in bulk PbTe \cite{Bajaj2015}. However, the determination of the minimum mean grain size is not easy, as small grains tend to anneal at finite temperatures until the single crystal is reached. Nevertheless, polycrystalline structures are found to be stable in experiments, and the reason is still unclear. This observation is not exclusive to PbTe, or to thermoelectric materials. Polycrystallinity is a very common phenomenon, and a full explanation for it is still lacking. Here we have shown that the presence of interfaces with reduced mobility (anisotropy) and the pinning pressure exerted by second particles (Zener effect), in this case voids, but it could be another second phase, e.g. as present in alloys, plays an essential role in the grain-growth stagnation \cite{Zener1949,Holm2010a}.

In our simulations, we observed that nanostructures of dimensions in the order of hundreds of nanometers could be stable at intermediate temperatures, as seen in different experiments \cite{Khasimsaheb2015,Kishimoto2015}, in the presence of voids as small as $20-50$ nm. The fabrication of nanostructured PbTe containing voids with these sizes (porous polycrystalline PbTe) is experimentally feasible \cite{Zimin2011}.  Small void sizes and high void fractions allow for the existence of small grains, as described by the limiting grain size given by equation \ref{eq:limitinggrainsize}.  This study sheds light into the general question of the metastability of polycrystalline samples against the single crystal. We have shown that grain growth is arrested by voids pinned at grain boundaries, with the size of voids and grains being determined by vacancy concentration and temperature. A similar phenomenon may be induced by impurities, instead of, or in addition to voids \cite{Humphreys2004}.

\section{Conclusions}\label{sec:conclusions}
The effect of Zener pinning by voids on the main grain size is investigated by computer simulations using the phase-field method. The main conclusions can be summarized as follows:
\begin{itemize}
\item In the presence of voids, the mean grain size follows approximately a log-normal distribution during the steady-state regime. This is in contrast with Hillert's distribution observed  in normal grain growth. This deviation in the distribution is likely to be due to the break down of scale-invariance shown in Fig. \ref{fig:expgrainwvoid300K}.
\item Grain growth in polycrystalline PbTe can be stopped by immobile, spherical voids until the mean grain size reaches a limiting value (see Fig. \ref{fig:limitinggrainsize}). Consequently, polycrystalline materials can reach stable structures and can be used to reduce the lattice thermal conductivity, as the latter drop for decreasing grain size.
\item The limiting grain size is proportional to the void radius at constant void fraction (see Eq. \ref{eq:limitinggrainsize}). Therefore, small void sizes are desired to stabilize samples with small mean grain sizes, and hence low thermal conductivity. Additionally, voids can reduce the thermal conductivity by themselves, so the presence of grain boundaries and voids in thermoelectric devices results in efficiency enhancement. 
\item The metastable void size, $r_v$, and hence the mean grain size, $R_Z$, grows with temperature (See inset to Fig. \ref{fig:metastablevoid}). Therefore, due to this coarsening phenomenon, also thermal conductivity should increase with temperature, thus casting doubts, in general, in the effectiveness of nanostructuring as a strategy for improving the thermoelectric figure of merit. Since the growth process is irreversible, the void size is determined by the coarsening temperature, i.e. the maximum temperature to which it has been exposed. PbTe, in particular, is a leading thermoelectric material at intermediate temperatures and is not usually exposed to high temperatures, so PbTe structures with small voids and a relatively small mean grain size can remain stable under operational conditions.
\item The lattice thermal conductivity of porous polycrystalline PbTe can be reduced by 35\% by anisotropic grain boundaries pinned by metastable voids (see Fig. \ref{fig:finalcondreduction}). The reduction in conductivity is larger when the sample has not first been exposed to higher temperatures.  
\end{itemize}

\ack
This work was supported by a research grant from Science Foundation 
Ireland (SFI) and the Department for the Economy Northern Ireland under 
the SFI-DfE Investigators Programme Partnership, Grant Number 15/IA/3160. We are grateful for computational support from the UK national high-performance computing service, ARCHER, for which access was obtained via the UKCP consortium and funded by EPSRC grant ref EP/P022561/1, and from the UK Materials and Molecular Modelling Hub, which was partially funded by EPSRC grant ref EP/P020194/1. We thank Adrian P. Sutton, Michael W. Finnis and Lorenzo Stella for insightful discussions.

\section*{References}
\providecommand{\newblock}{}


\end{document}